\documentclass[11pt,a4paper,aps]{revtex4}
\usepackage{graphicx}
\def\be{\begin{equation}}
\def\ee{\end{equation}}
\def\bea{\begin{eqnarray}}
\def\eea{\end{eqnarray}}
\def\no{\nonumber}
\def\bef{\begin{figure}}
\def\ef{\end{figure}}
\def\bc{\begin{center}}
\def\ec{\end{center}}
\def\wt{\widetilde}
\begin{document}
\title{Generalized coherent states of exceptional Scarf-I potential: Their spatio-temporal and statistical properties}

\author{T. Shreecharan} 
\email{shreecharan@ifheindia.org}
\affiliation{Department of Physics, Faculty of Science and Technology,\\
ICFAI Foundation for Higher Education, \\
Dontanapally, Hyderabad, Telangana, India: 501 203.}

\author{S. Sree Ranjani }
\email{s.sreeranjani@ifheindia.org }
\affiliation{Department of Physics, Faculty of Science and Technology,\\
ICFAI Foundation for Higher Education, \\
Dontanapally, Hyderabad, Telangana, India: 501 203.}
\begin{abstract}
\begin{center}
\textbf{\small{Abstract}}
\end{center}
We construct generalized coherent states for the rationally extended Scarf-I potential. Statistical and geometrical properties of these states are investigated. Special emphasis is given to the study of spatio-temporal properties of the coherent states via the quantum carpet structure and the auto-correlation function. Through this study, we aim to find the signature of the ``rationalisation" of the conventional potentials and the classical orthogonal polynomials.
\end{abstract}
\maketitle

\section{Introduction}

The discovery of exceptional orthogonal polynomials (EOPs) \cite{Kamran1,Kamran2} has had a significant impact on the areas of Sturm-Liouville theory and exactly solvable quantum mechanics \cite{Sasaki1}. The latter has seen a revival, as the existence of these polynomials has led to the contruction of new exactly solvable potentials \cite{Quesne1,Sasaki2}. Many groups worked in tandem using various methods such as Darboux transformations, Darboux-Crum, supersymmetry (SUSY), group theory etc., to construct these new polynomials and potentials
\cite{kam_dar,Sasaki3,Dutta,Ramos,Grandati,Mandal,Chaitanya,Sreeranjani1}.  Another interesting development that has occurred recently is the construction of multi-indexed EOPs leading to generalised families of new potentials  \cite{Sasaki4,Quesne2,multi1,multi2,Grandati-Quesne,Sreeranjani2}.

These potentials are rational extensions of the conventional exactly solvable potentials wherein additional rational terms get added, hence the new potentials are also called as the \textit{rational potentials}. Interestingly the old and the corresponding new potentials are isospectral to each other. While the wavefunctions of the rational potentials involve EOPs, the conventional potentials have the classical orthogonal polynomials (COPs) as their eigenfunctions. Like the potentials, the EOPs are also rational extensions of the COPs.

Various aspects of the conventional potentials have been extensively explored. One such aspect is the construction of the coherent states (CS). Ever since the introduction of coherent states, as the most classically behaving quantum states, by Schr\"{o}dinger, this concept has evolved over the years. At present there are three universally accepted definitions of CS. The first is that these states be the eigenstates of the annihilation operator. The second definition states that the coherent state must be generated by a unitary displacement operator. The third definition is that the CS must saturate the uncertainty relation. It is well known all three definitions coincide only for the harmonic oscillator potential, which obeys the Heisenberg-Weyl algebra. Other exactly solvable potentials have either the $su(2) $ or the $su(1,1)$ algebra as their underlying algebra, therefore the CS corresponding to a given potential are distinct and do not coincide. This makes the study of these states and their properties quite fruitful and exciting.

Though the Lie algebraic techniques are very powerful and elegant, it is sometimes quite difficult to establish the generators of the algebra in terms of the potential parameters. This makes the construction of the CS difficult and therefore we have to resort to other methods. In this context the  shape invariance symmetry or the discrete reparametrisation symmetry of these potentials has played a pivotal role \cite{Gedenstein,Aizawa1,Balantekin1,Balantekin2}. Since the rational potentials also respect reparametrisation symmetry, it is natural that we exploit this to construct the CS. Similar studies for the new rational potentials and the underlying dynamical symmetry groups have been done \cite{Aizawa2}. Recently CS for the rationally extended oscillator have been constructed and their statistical and dynamical properties have been studied \cite{hoffman1,hoffman2,hoffman3}. 

We construct the generalised CS (GCS) as defined in Ref. \cite{Balantekin2}, for the rational Scarf-I potential, which has $X_1$ EOPs as solutions. In addition to studying their statistical and geometrical aspects, we also study the quantum carpet structures.  These are interesting patterns generated by the spatio-temporal evolution of  the probability density function of the CS. We also study the auto-correlation functions of the rational CS. All the results obtained for the rational Scarf-I potentials are compared with those of the conventional Scarf-I potential, so as to obtain the signature of rationalization of the potential and the wave functions.

The paper is organized as follows. In section II, we summarize the rational and the conventional Scarf-I potentials and their solutions and briefly summarise supersymmetry (SUSY) and shape invariance (SI). In section III, we present the four types of GCS for this potential. This is followed by the quantum carpets structures of the old and the new potentials and their comparisons in section IV. Section V, deals with the auto-correlation function, while section VI deals with the statistical and geometrical aspects of the GCS. In section VII, we present our conclusions.

\section{The rational trigonometric Scarf-I potential}


As is well known the shape invariant potentials (SIPs) are classified into six categories based on the Infeld and Hull factorization scheme \cite{Infeld}. This classification is based on the translational transformation properties of the potential parameters. The trigonometric Scarf-I potential belongs to the type $\mathcal{A}$ of  SIPs as per this scheme. The rational Scarf-I has the same reparametrisation symmetry as the conventional potential therefore it can also be classified as a type $\mathcal{A}$ potential.

In this section, we present the Scarf-I and the rational Scarf-I potentials and their solutions. These will be used in our subsequent construction of GCS and their properties. The conventional supersymmetric trigonometric Scarf-I potential  with $(\hbar=2m=1)$ is given by
\begin{equation} \label{potcon}
V^-(x) = [\alpha(\alpha-1)+\beta^2]\sec^2x  - \beta(2\alpha-1)\sec x \tan x ,
\end{equation}
whose eigenfunctions and eigenvalues are
\begin{equation} \label{wfcon}
 \psi^-_{n}(x) = N_n (1-\sin x)^{\frac{(\alpha-\beta)}{2}}(1+\sin x)^{\frac{(\alpha+\beta)}{2}}  P_{n}^{(\alpha-\beta-\frac{1}{2}\,,\,\alpha+\beta-\frac{1}{2})}(\sin x),
\end{equation}
and
\begin{equation}
E^-_{n}= (n+\alpha)^2, \qquad  n=0,1,2,\dots
\end{equation}
respectively. Here $P_{n}^{(\alpha-\beta-\frac{1}{2}\,,\,\alpha+\beta-\frac{1}{2})}(\sin x)$
are the classical Jacobi polynomials and the normalization constant in Eq. (\ref{wfcon}) is
\begin{equation}
\hskip-0.5cm N_{n}=\sqrt{\frac{n!(2n+\alpha)\Gamma(n+2\alpha)}{2^{2\alpha}\Gamma(\alpha-\beta+n+\frac{1}{2})\Gamma(\alpha+\beta+n+\frac{1}{2})}}    \label{Ncon}
\end{equation}
The corresponding superpotential associated with this potential is given by
\begin{equation}
W(x) = \alpha \tan x - \beta \sec x \ .
\end{equation}


\noindent The rational trigonometric Scarf-I potential is given by
\begin{eqnarray}\nonumber
\wt{V}^-(x) & = & [\alpha(\alpha-1)+\beta^2]\sec^2x - \beta(2\alpha-1)\sec x \tan x \\
& + & \frac{2(2\alpha-1)}{2\alpha-1-2\beta \sin x} -  \frac{2[(2\alpha-1)^2-4\beta^2]}{(2\alpha-1-2\beta \sin x)^2},\label{pot}
\end{eqnarray}
where $-\frac{\pi}{2} <x<\frac{\pi}{2}$ and $0<\beta<\alpha-1$. The  bound state solutions are
\begin{equation} \label{wf1}
 \wt{\psi}^-_{n}(x)= N_{n}\frac{(1-\sin x)^{\frac{1}{2}(\alpha-\beta)}(1+\sin x)^{\frac{1}{2}(\alpha+\beta)}}{2\alpha-1-2\beta \sin x} \\  \wt{P}_{n+1}^{(\alpha-\beta-\frac{1}{2},\alpha+\beta-\frac{1}{2})}(\sin x) \ ,
\end{equation}
where
\begin{eqnarray} \no
\wt{P}_{n+1}^{(\alpha-\beta-\frac{1}{2},\alpha+\beta-\frac{1}{2})}(\sin x) = -\frac{1}{2}\left[\sin x-\frac{2\alpha-1}{2\beta}\right] P_{n}^{(\alpha-\beta-\frac{1}{2},\alpha+\beta-\frac{1}{2})}(\sin x) \\ \no
+ \frac{1}{2\alpha-1+2n}\Bigg[\frac{2\alpha-1}{2\beta} P_{n}^{(\alpha-\beta-\frac{1}{2},\alpha+\beta-\frac{1}{2})}(\sin x) -P_{n-1}^{(\alpha-\beta-\frac{1}{2},\alpha+\beta-\frac{1}{2})}(\sin x)\Bigg]
\end{eqnarray}
are the $X_1$ exceptional Jacobi polynomials and
\begin{eqnarray} \no
\wt{N}_{n}= \frac{\beta}{2^{\alpha-2} } \Bigg[\frac{n !(2n+2\alpha)\Gamma(n+2\alpha)}{(n+\alpha-\beta+\frac{1}{2})(n+\alpha+\beta+\frac{1}{2})}
\\ \label{norm1}
\frac{1}{\Gamma(n+\alpha-\beta- \frac{1}{2})\Gamma(n+\alpha+\beta-\frac{1}{2})}\Bigg]^\frac{1}{2}
\end{eqnarray}
is the normalization constant. The corresponding energy eigenfunctions are given by
\begin{equation}
\wt{E}^-_{n}= (n+\alpha)^2, \quad \quad  n=0,1,2,\dots     \label{en1}
\end{equation}
The superpotential associated with this potential is given by
\begin{eqnarray} \no
W(x)=\alpha \tan x - \beta \sec x \\ \label{suppot}
- 2 \beta \cos x \Bigg[\frac{1}{2\alpha-1-2\beta \sin x}-  \frac{1}{2\alpha+1-2\beta \sin x} \Bigg] .
\end{eqnarray}
It is clear from the above expressions, as to why the new potentials and the polynomials are called the rational extensions of their conventional and classical counterparts respectively. For both the supersymmetric potentials we can construct their isospectral partners $V^+(x)$ and $\wt{V}^+(x)$, using the given superpotentials respectively. Similar to the Scarf-I, the rational Scarf-I potential  satisfies the shape invariance condition
\begin{equation}
\wt{V}^+(x,a_1)=\wt{V}^-(x, a_2)+R(a_1),
\end{equation}
where $R$ is a function of $a_1$ with $a_1=\alpha$ and $a_2=\alpha+1$ for the Scarf-I potential. For further details on supersymmetric quantum mechanics, we refer to \cite{Khare-book}.

\begin{figure}[hbt]
\centering
\includegraphics[height=2in,width=2.5in]{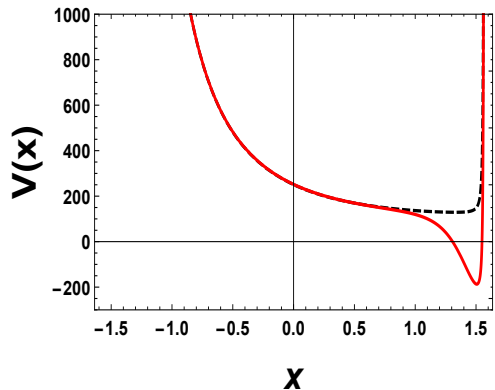}
\caption{The potential with $\alpha =12$ and $\beta = 10.9$. The dashed line is for the Scarf-I potential and the continuous curve depicts the rational Scarf-I potential.}\label{figure1}
\end{figure}

In figure (\ref{figure1}), we give the plots of these potentials for certain parameter values.
As can be seen, the rational terms in the potential lead to a small dip at one end of the potential well. As will be seen later in the paper this dip has an effect on the quantum carpet structure.

\section{The coherent state construction}

For any $j$th potential in the hierarchy of SIPs, the GCS, as defined in Ref. \cite{Balantekin2}, is
\begin{equation} \label{csbalantekin}
\big|\zeta; a_j \big\rangle = \frac{1}{\sqrt{N \big(|\zeta|^2;a_j\big)}} \sum_{n=0}^\infty \frac{\zeta^n}{h_n(a_j)} \ \big|\psi_n \big\rangle \ .
\end{equation}
In the above $h_n(a_j)$ are the expansion coefficients defined below
\begin{equation} \label{weights}
h_n(a_j) = \sqrt{\prod_{k=1}^{n}\Big[\sum_{s=k}^{n}R(a_s)\Big]} \Bigg/\prod_{k=1}^{n-1} \mathcal{Z}_{j+k}
\end{equation}
for  $n \geq 1$. We have used the notation  \\ 
$(a_j) \equiv [R(a_1), R(a_2), R(a_3), \cdots, R(a_n); a_j, a_{j+1}, \cdots a_{j+n-1}]$ and $h_0 (a_j) = 1$. The index $j$ is the family index in the hierarchy of shape invariant potentials. These potentials are related to each other by a translational shift of the potential parameter.The normalization constant is given by
\begin{equation}
N\big(|\zeta|^2;a_j\big) = 1 \Bigg/ \sqrt{\sum_{n=0}^{\infty}\frac{|\zeta|^{2n}}{|h_n(a_j)|^2}} \ .
\end{equation}
Note that $\big|\psi_n \big\rangle $ is the eigenstate of the SIP and the eigenvalues are given by $E_n = \sum_{k= 1}^{n} R(a_k)$ \cite{Khare-book}.
In the above equation  $\mathcal{Z}_{j+k} \equiv \mathcal{Z}(a_1, a_2, \cdots, a_{j+k})$ is an arbitrary functional of the potential parameters. It plays a crucial role in the construction of the CS. It is worth mentioning that $Z_j$ is not completely arbitrary, but is constrained by the shape invariance condition. $\mathcal{Z}_{j+k}$ and $\mathcal{Z}_{j}$ are related via a similarity transformation which is given by $\mathcal{Z}_{j+k}= \{\hat{T}(a_1)\}^k \mathcal{Z}_{j} \{\hat{T}^\dagger (a_1)\}^k$. 

A few comments are in order here about the operator $\hat{T}(a_1)$: (1) It is unitary (2) Its action on an eigenstate $|\psi_n (a_j) \rangle$ corresponding to a SIP is given by $\hat{T}(a_1) |\psi_n (a_{j+1}) \rangle$ (3) Its action on an operator $O(a_j)$ is via a similarity transformation $\hat{T}(a_1) \ O(a_{j}) \ \hat{T}^\dagger (a_1) = O(a_{j+1}) $. For more details about the operator $\hat{T}(a_1)$ and the details of the construction of CS, the readers should see Ref \cite{Aizawa1,Balantekin1,Balantekin2}.

Four different types of CS are constructed, for the Scarf-I potential and its rational extension, by suitably defining $Z_j$ in terms of an auxiliary function $g(c,d,a_j)$. Here $c$ and $d$ are constants and $a_1=\alpha, a_2= \alpha+1, \cdots, a_j = \alpha+j-1$. Below we present the expressions for the four different CS. Note in what follows for the sake of notational simplicity we suppress the $a_j$ dependency in the expressions of the CS.

\vskip0.3cm
\noindent\textbf{GCS 1:}
\vskip0.3cm

\noindent $\mathcal{Z}_{j} = C$ (constant) leads to the coherent state constructed by Aizawa et. al. \cite{Aizawa2}
\bea
\big|\zeta\big \rangle = \frac{1}{\sqrt{N \big(|\zeta|^2 \big)}}\sum_{n=0}^\infty \sqrt{\frac{\Gamma(2\alpha+n) }{\Gamma(n+1)\Gamma(2\alpha+2n)}} \  \zeta^{n} \ \big|\psi_n\big\rangle \ .
\eea
Here,
\be
\frac{1}{\sqrt{N \big(|\zeta|^2\big)}} = \sum_{n=0}^\infty \frac{\Gamma(2\alpha+n) |\zeta|^{2n}}{\Gamma(n+1)\Gamma(2\alpha+2n)}  ,
\ee
which can be summed to obtain
\be \label{norm-gcs1}
N(|\zeta|^2) ={ _1}F_2 \big( 2 \alpha; \alpha, \alpha + \frac{1}{2}; \frac{|\zeta|^2}{4}\big) \ .
\ee
For the other three CS the auxiliary function $g(c,d,a_j)$ is non trivial and of the form
\be \label{g}
g(a_j;c,d)= c a_j+d    ,
\ee
such that
\be \label{pi_g}
\prod _{k=0}^{n-1} g(a_{j+k};c,d)=\frac{c^n \Gamma(\alpha +j+n+\frac{d}{c}-1)}{\Gamma(\alpha +j+\frac{d}{c}-1)}
\ee

\vskip0.3cm
\noindent\textbf{GCS 2:}
\vskip0.3cm

\noindent By defining
\be
\mathcal{Z}_{1} = \sqrt{g(\alpha; 2,1) \ g(\alpha; 2, 2)} \ e^{-i \tilde{\alpha} R(a_1)}
\ee
and using Eq. (\ref{pi_g}), we are led to the second type of GCS
\bea \no
\big|\zeta\big \rangle = \frac{1}{\sqrt{N \big(|\zeta|^2\big)}} \sum_{n=0}^{\infty} \sqrt{\frac{\Gamma(2\alpha+n) \Gamma(2\alpha+2n+1)}{\Gamma(n+1)\Gamma(2\alpha+2n) \Gamma(2\alpha+1)}} \\ \label{2cs}
e^{-i\tilde{\alpha}E_n} \zeta^n \big| \psi_n \big\rangle ,
\eea
which is the Perelomov coherent state for the Scarf-I potential. Here $\tilde{\alpha}$ is a real constant. The normalization constant for this case turns out to be
\be \label{norm-gcs2}
N \big(|\zeta|^2\big) = {_2}F_1 \big( 2 \alpha; \alpha +1, \alpha; |\zeta|^2\big) \ .
\ee

\vskip0.3cm
\noindent\textbf{GCS 3:}
\vskip0.3cm

\noindent The third GCS, which happens to be a Barut-Girardello coherent state, is obtained by defining
\be
\mathcal{Z}_{1} = \frac{\sqrt{g(\alpha; 2, 1) g(\alpha; 2, 2)}}{g(\alpha; 1,1+ \alpha)} \ e^{-i \tilde{\alpha} R(a_1)} \ .
\ee
The expression for the coherent state turns out to be
\bea 
\big|\zeta \big \rangle = \frac{1}{\sqrt{N \big(|\zeta|^2)}} 
\sum_{n=0}^{\infty}\sqrt{\frac{\Gamma(2\alpha+2n+1) \Gamma(2\alpha+n) \Gamma(2\alpha+1)}{\Gamma(2\alpha+2n) \Gamma(2\alpha+n+2) \Gamma(n+1)}} \ e^{-i\tilde{\alpha}E_n} \zeta^n \big| \psi_n \big\rangle ,
\eea
where
\bea \label{norm-gcs3}
N \big(|\zeta|^2\big) = \frac{\Gamma(2\alpha+1)^2}{\Gamma(2\alpha+2)}\ {_2} F_2 \big( 2 \alpha, \alpha+1; \alpha, 2\alpha + 2; |\zeta|^2\big) \ .
\eea

\vskip0.3cm
\noindent\textbf{GCS 4:}
\vskip0.3cm

\noindent The fourth coherent state is obtained by defining
\bea
\mathcal{Z}_{j} = \sqrt{\frac{g(a_j; 1, 0) \, g(a_j; 1, 1/2) \, g(a_{j+2}; 2, -2\alpha-2\sigma)}{g(a_j;1,\alpha)\, g(a_{j+2}; 1, -\alpha)}} \ e^{-i\tilde{\alpha} R(a_1)} .
\eea
Setting $j=1$ in the above equation we get
\bea
\mathcal{Z}_{1} = \sqrt{\frac{g(\alpha; 1, 0) \, g(\alpha; 1, 1/2) \, g(\alpha+2; 2, -2\alpha-2\sigma)}{g(\alpha;1,\alpha)\, g(\alpha+2; 1, -\alpha)}} \ e^{-i\tilde{\alpha} R(a_1)} ,
\eea
which leads to
\bea
\big|\zeta \big \rangle =  \frac{1}{\sqrt{N \big(|\zeta|^2 \big)}} \sum_{n=0}^\infty\sqrt{\frac{\Gamma(n+2-\sigma)}{\Gamma(n+2)\Gamma(n+1)}} \ e^{-i\tilde{\alpha} E_n}\zeta^n \big|\psi_n\big \rangle ,
\eea
where $\sigma$ is a real number dependent on  $d$ via equations (\ref{g})  and (\ref{pi_g}), whose explicit form is $\sigma = - \alpha -d/2$. The normalization constant is
\be \label{norm-gcs4}
N \big(|\zeta|^2\big) = \Gamma(2-\sigma) \ {_1}F_1(2-\sigma; 2; |\zeta|^2) \ .
\ee
As already mentioned $\big|\psi_n\big \rangle$ are the normalized wave functions, Eq. (\ref{wfcon}), of the Scarf-I potential. Replacing these by $\wt{\psi}_n(x)$ from Eq. (\ref{wf1}) in the above expressions, will yield the corresponding GCS for the rational potential. 

\section{Quantum Carpet Structure}

In the section we study the spatio-temporal behaviour of the GCS of the conventional Scarf-I potential and its rational extension. By evolving the probability density of the CS over space-time leads to quantum carpet structures \cite{Berry,Rost,Marzoli,Loinaz,Hall,Shree}. These patterns are generated because of the revival dynamics exhibited by the wavepacket. This idea of the revival dynamics was first put forth in the work of Ref. \cite{Averbukh}. A revival is said to take place when a wavepacket under time evolution returns to its initial form. Fractional revival is said to occur when a time evolving wavepacket breaks up into sub-packets at different locations, but with a shape that closely matches the initial form. Since the eigenvalues of the Scarf-I and the rational Scarf-I potential are quadratic, there will be revivals as well as fractional revivals. Potentials that have linear energy spectrum do not exhibit such a behaviour. For mathematical details concerning the revival dynamics we refer to the exhaustive review article of Robinett \cite{Robinettphysrep}.


\begin{figure}
\centering
\begin{minipage}{2.3in}
\centering
\includegraphics[height=2.2in,width=2.2in]{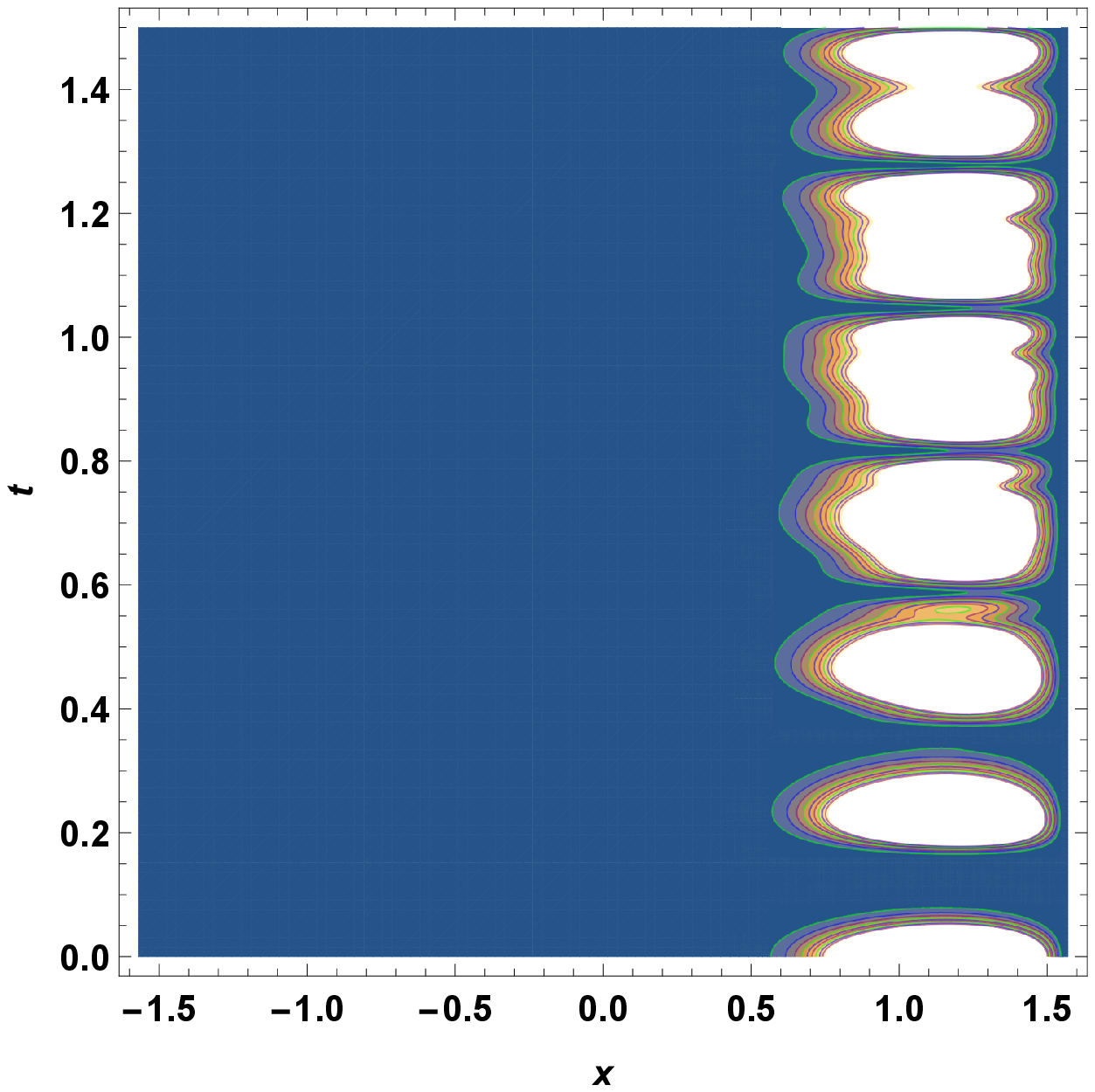}
\caption{Carpet structure for GCS1, Scarf-I.} \label{carpet-con1}
\end{minipage}\hfill
\begin{minipage}{2.3in}
\centering
\includegraphics[height=2.2in,width=2.2in]{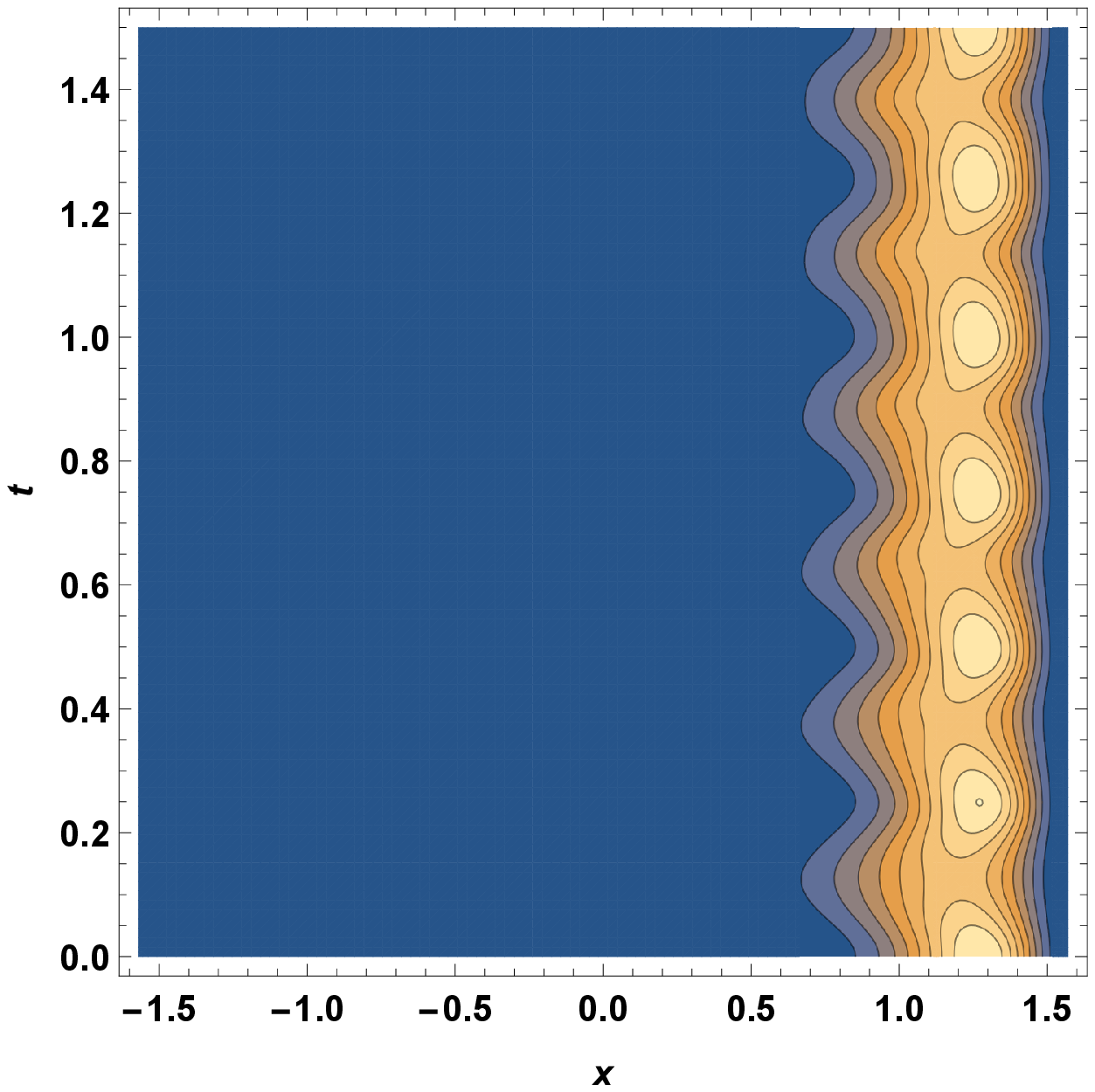}
\caption{Carpet structure for GCS1, rational Scarf-I.} \label{carpet-rat1}
\end{minipage}
\end{figure}


\bef
\centering
\begin{minipage}{2.3in}
\centering
\includegraphics[height=2.2in,width=2.2in]{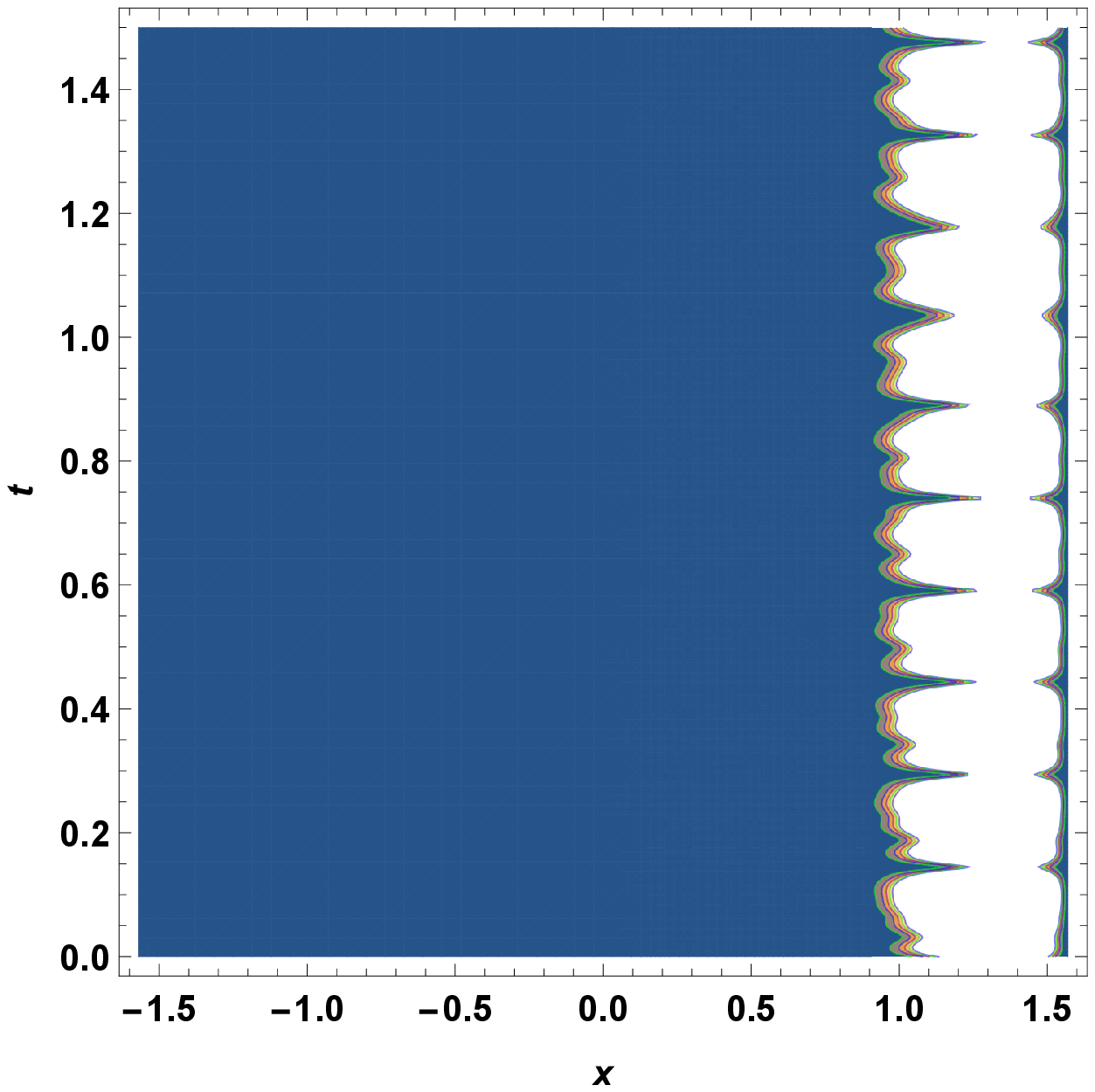}
\caption{Carpet structure for GCS2, Scarf-I.} \label{carpet-con2}
\end{minipage}\hfill
\begin{minipage}{2.3in}
\centering
\includegraphics[height=2.2in,width=2.2in]{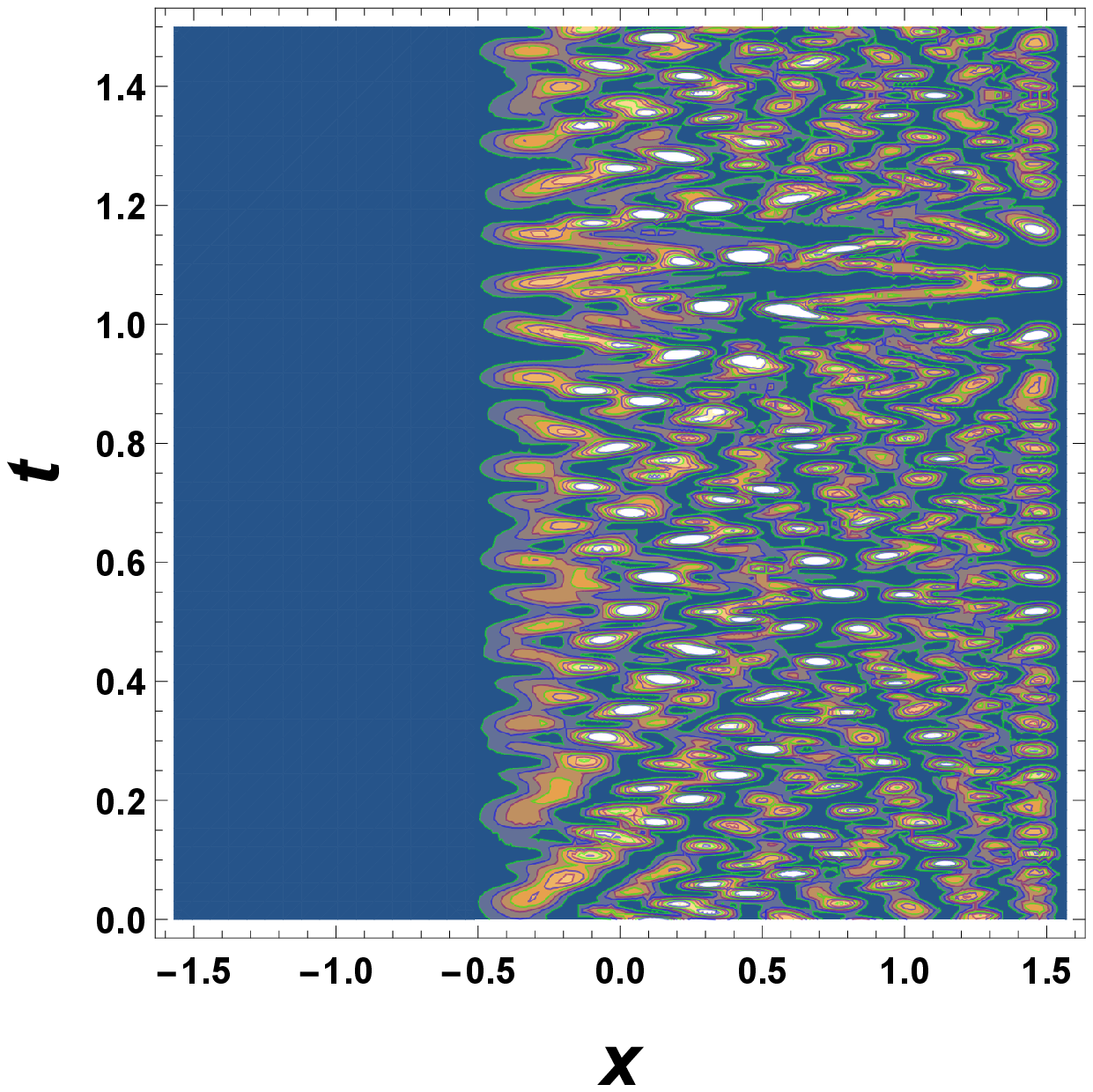}
\caption{Carpet structure for GCS2, rational Scarf-I.} \label{carpet-rat2}
\end{minipage}
\ef

\bef
\centering
\begin{minipage}{2.3in}
\centering
\includegraphics[height=2.2in,width=2.2in]{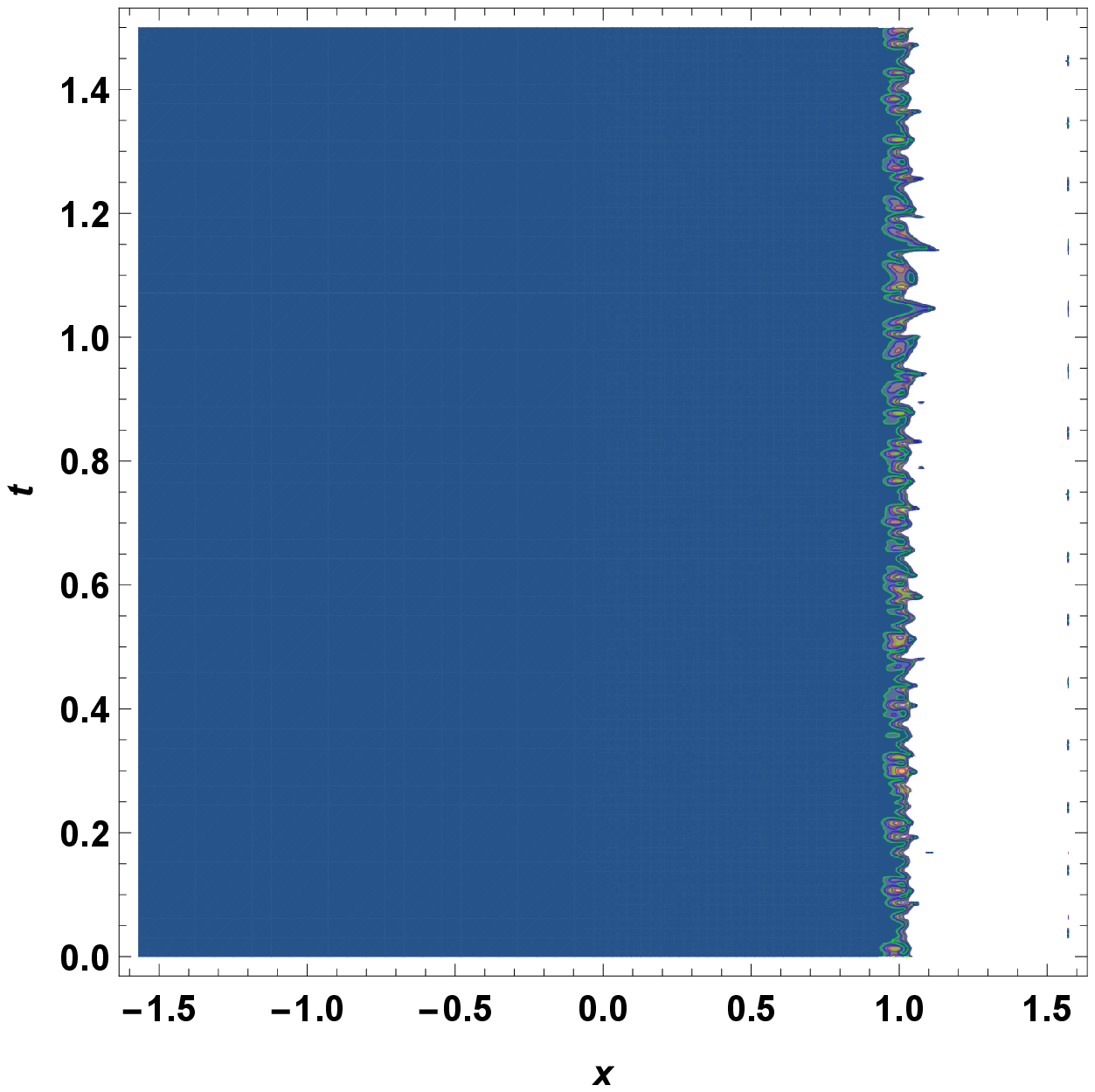}
\caption{Carpet structure for GCS3, Scarf-I.}\label{carpet-con3}
\end{minipage}\hfill
\begin{minipage}{2.3in}
\centering
\includegraphics[height=2.2in,width=2.2in]{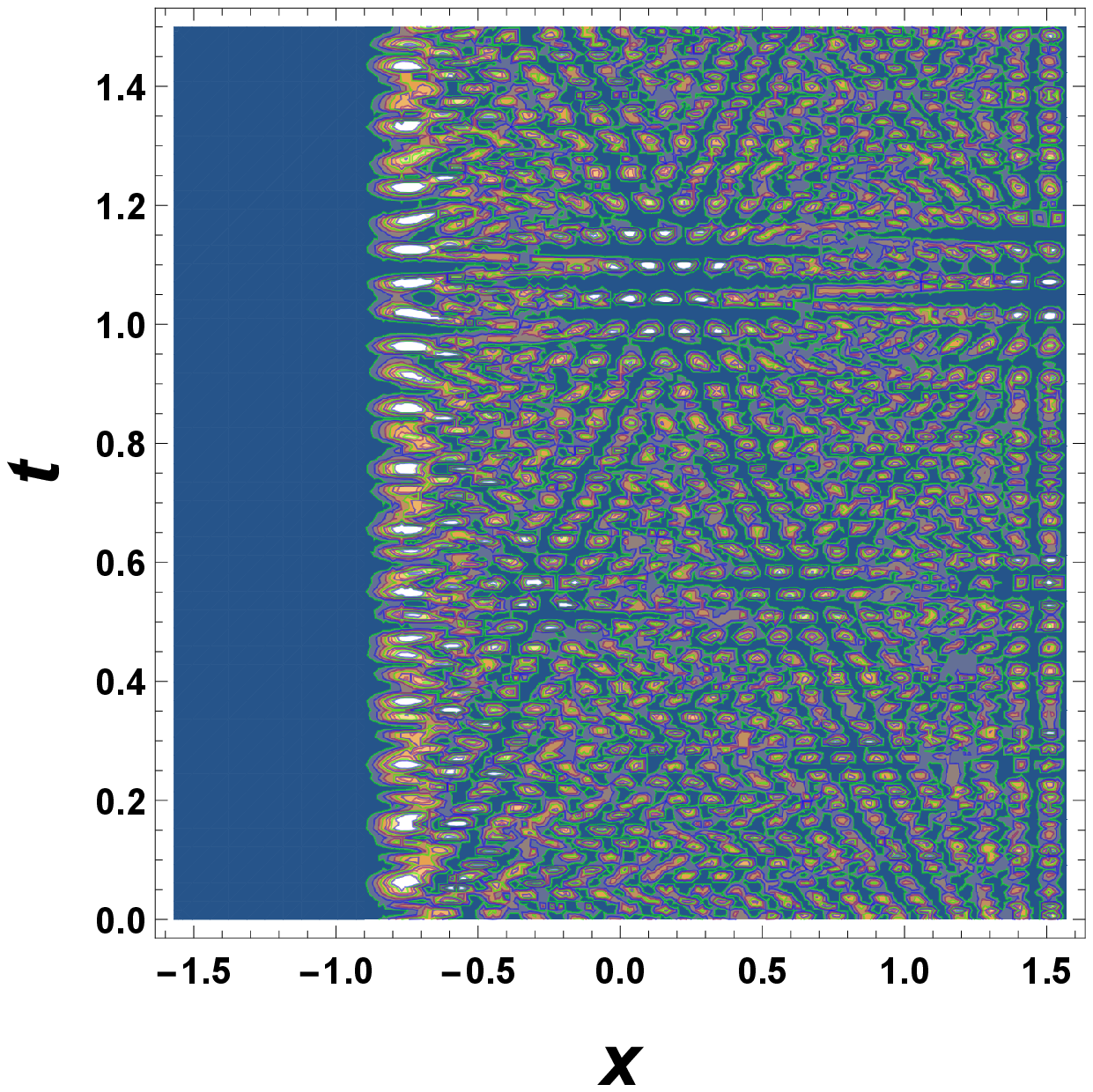}
\caption{Carpet structure for GCS3, rational Scarf-I.}\label{carpet-rat3}
\end{minipage}
\ef

\bef
\centering
\begin{minipage}{2.3in}
\centering
\includegraphics[height=2.2in,width=2.2in]{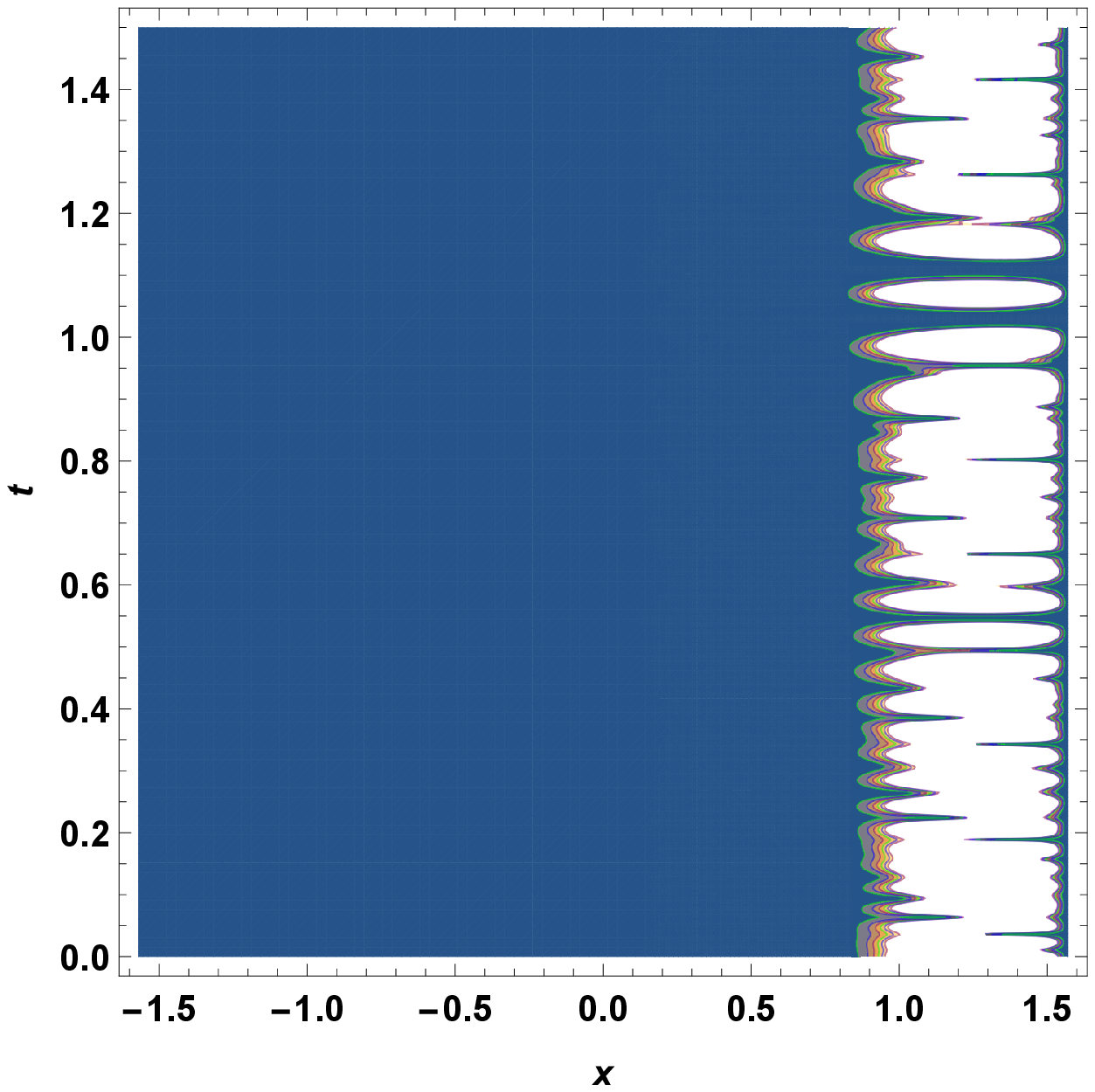}
\caption{Carpet structure for GCS4, Scarf-I.} \label{carpet-con4}
\end{minipage}\hfill
\begin{minipage}{2.3in}
\centering
\includegraphics[height=2.2in,width=2.2in]{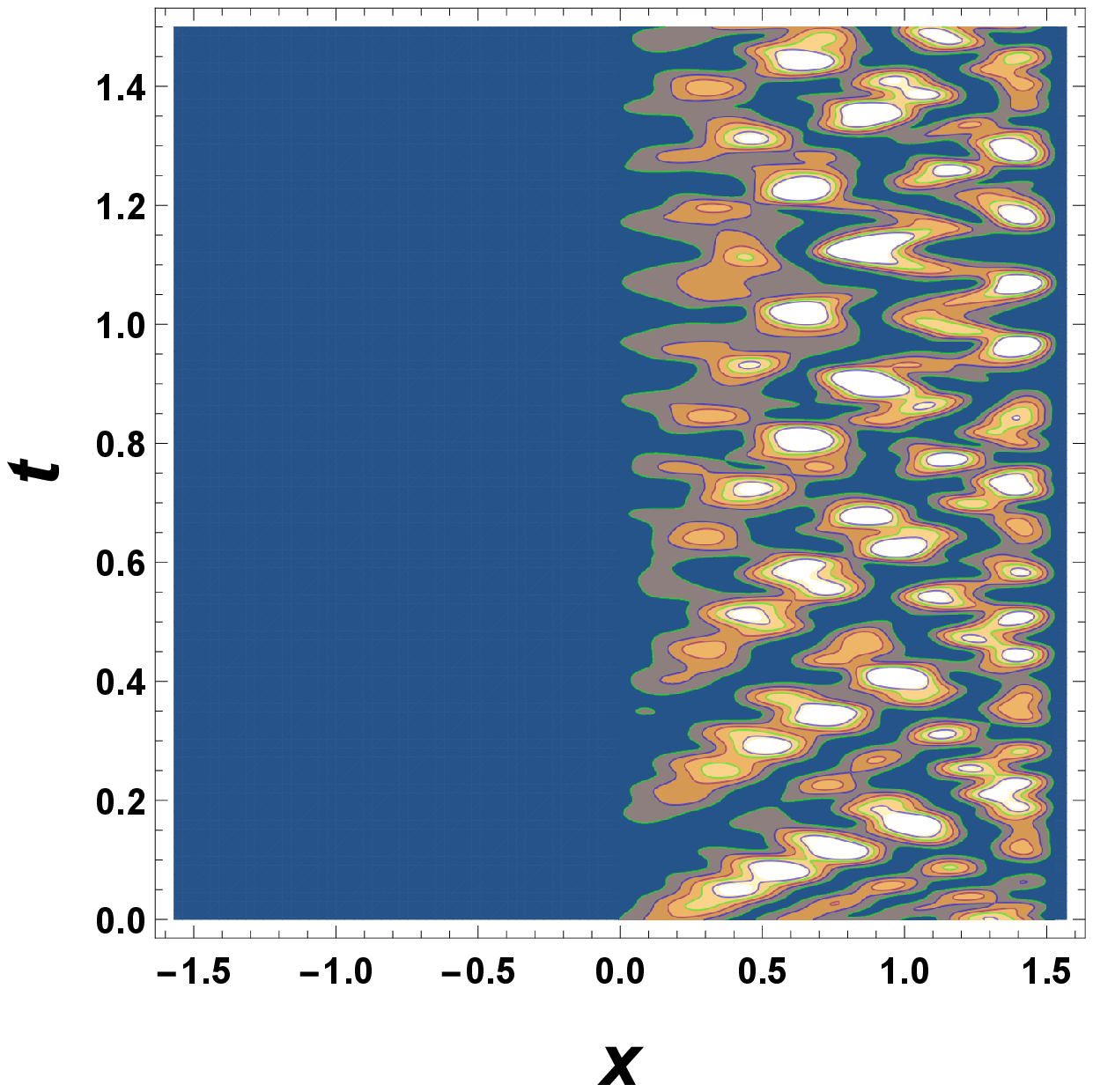}
\caption{Carpet structure for GCS4, rational Scarf-I.}\label{carpet-rat4}
\end{minipage}
\ef
%
%
The quantum carpet structures corresponding to the GCS (1-4), for the conventional Scarf-I Potential, are shown in Figures (\ref{carpet-con1}), (\ref{carpet-con2}), (\ref{carpet-con3}), and (\ref{carpet-con4}) respectively. In parallel, the four GCS for the rational Scarf-I potential lead to the quantum carpet structures that are shown in the Figures (\ref{carpet-rat1}), (\ref{carpet-rat2}), (\ref{carpet-rat3}), and (\ref{carpet-rat4}) respectively. The parameter values taken for the above carpets are $\alpha=12, \beta = 10.9$ and $n=20$. For the GCS4 case $\sigma = - \alpha - d/2$ with $d=0$.

In all the carpets the white regions indicate a very high probability of finding the particle as compared to other areas. From the plot of the potentials, it is clear that the wells are asymmetric about their center. This leads to the localisation of the particle to the right end of the well as seen in all the quantum carpet figures.

The additional terms in the rational Scarf-I potential lead to a dip near the right wall leading to an increase in the asymmetry as compared to the conventional potential. This has a profound effect on the quantum carpet structure of the rational potential. The high probability regions of the carpet which for the conventional potential were thick white bands, exhibits intricate substructures for the rational potential. 

These intricate patterns in our carpet structures, corresponding to the rational Scarf-I, are because of the wavepacket splitting caused by two reasons. First because of a dip due to the rational terms in the potential and the second due to the the revival and fractional revival phenomena occurring as a result of the quadratic energy structure. The dip leads to increased wave packet collisions not only with the walls of the potential but also with each other. These revival aspects are studied in the next section using the auto-correlation function. It is worth mentioning that in Ref. \cite{hoffman1} the authors have shown a similar wavepacket breaking in the rationally extended oscillator potential due to the dip in the center that this potential has as compared to the usual parabolic potential of an ordinary oscillator.

\section{Auto-correlation function/Fidelity}

The autocorrelation function or fidelity is defined as the overlap between the state at $t=0$ and at a later time $t$:
\begin{equation}
A(t) = \big\langle \zeta; a_r; t\big|\zeta; a_r; 0 \big\rangle \ .
\end{equation}
This quantity provides further evidence of the revival dynamics. Numerically $|A(t)|^2$ lies between 0 and 1. Zero means no overlap while one indicates a perfect overlap, pointing to a complete revival of the CS. The fidelity is very sensitive to the weighting factors of the CS. This is evident from the different crest and trough structure seen in the plots of $|A(t)|^2$ of all the four GCS. These are plotted in the Figures (\ref{fidelityrat1}), (\ref{fidelityrat2}), (\ref{fidelityrat3}), and (\ref{fidelityrat1}) corresponding to the GCS 1, 2, 3, and 4 respectively of the the rational Scarf-I potential. It must be pointed out that the autocorrelation plots for the Scarf-I potential will be the same.

\bef
\centering
\begin{minipage}{2.3in}
\centering
\includegraphics[height=2in,width=2.2in]{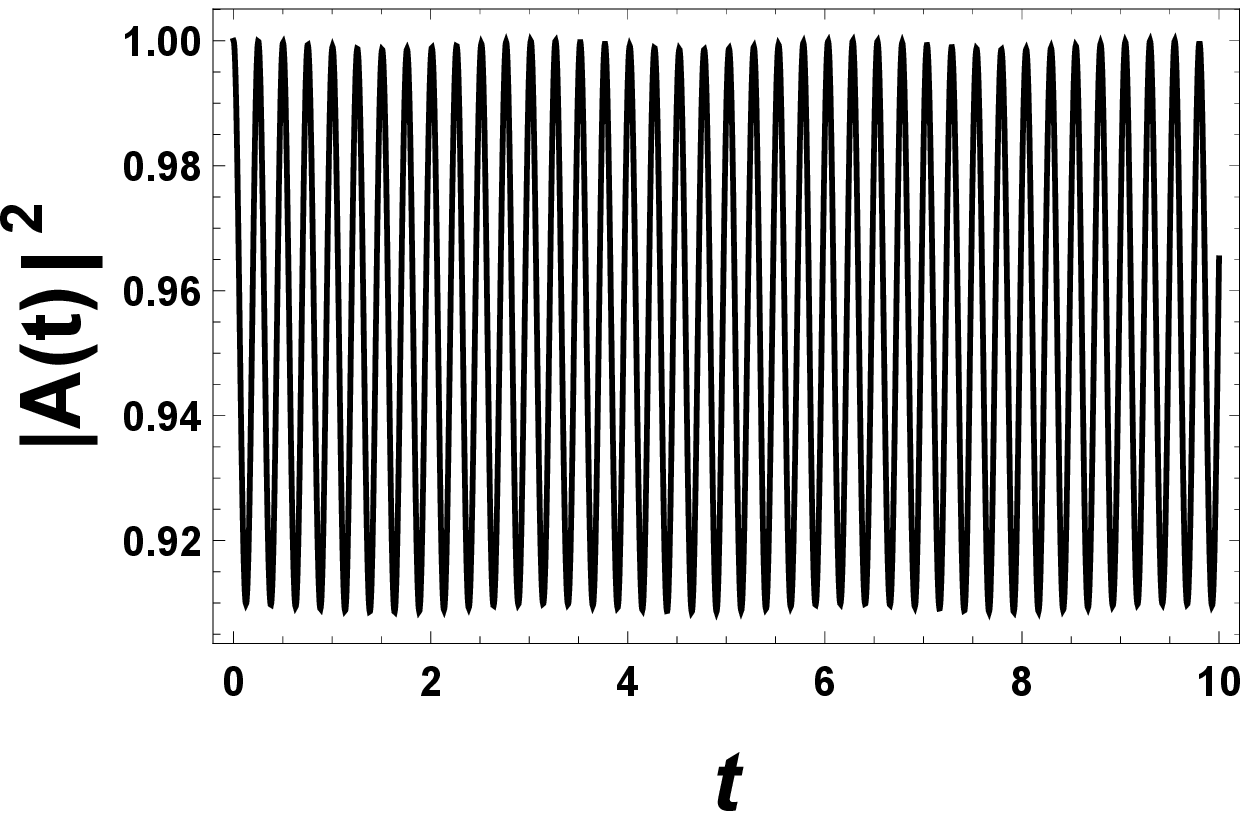}
\caption{Auto correlation function for GCS1.} \label{fidelityrat1}
\end{minipage}\hfill
\begin{minipage}{2.3in}
\centering
\includegraphics[height=2in,width=2.2in]{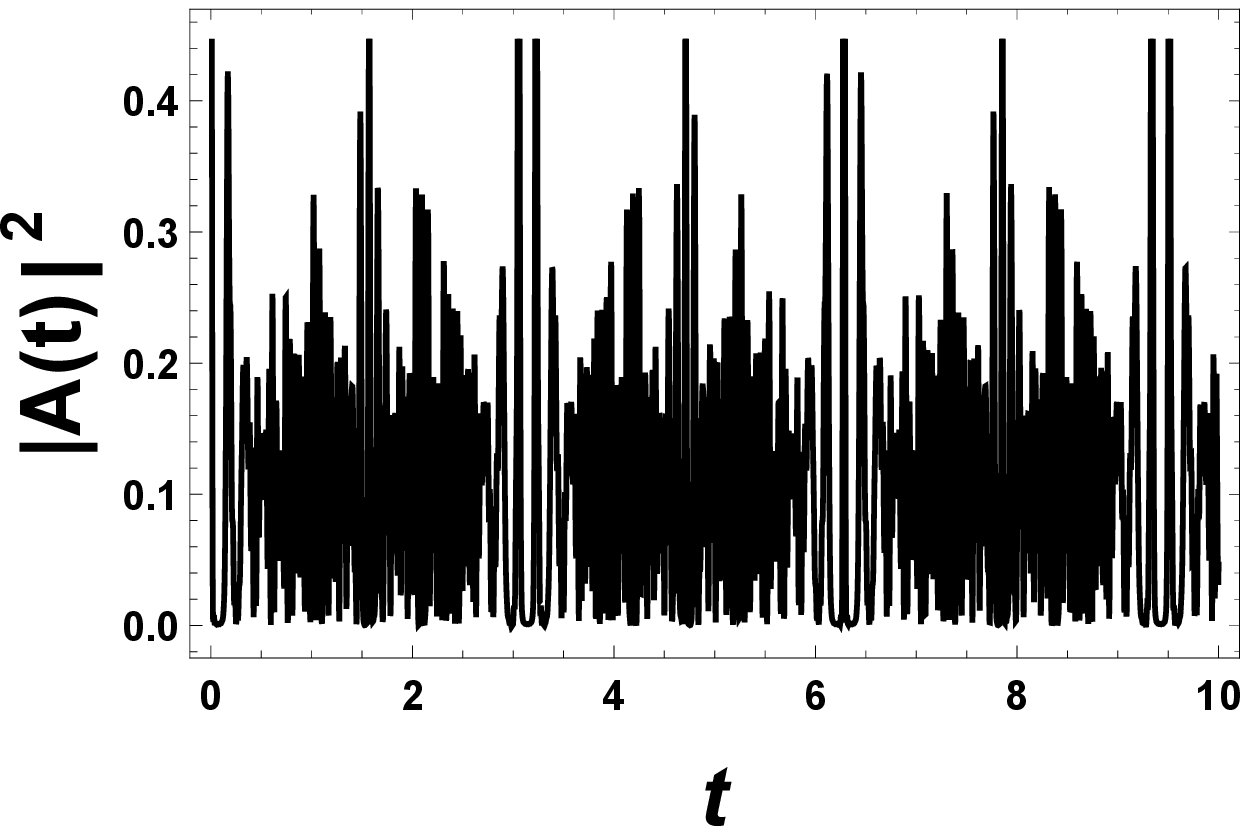}
\caption{Auto correlation function for GCS2.}\label{fidelityrat2}
\end{minipage}
\ef
\bef
\centering
\begin{minipage}{2.3in}
\centering
\includegraphics[height=2in,width=2.2in]{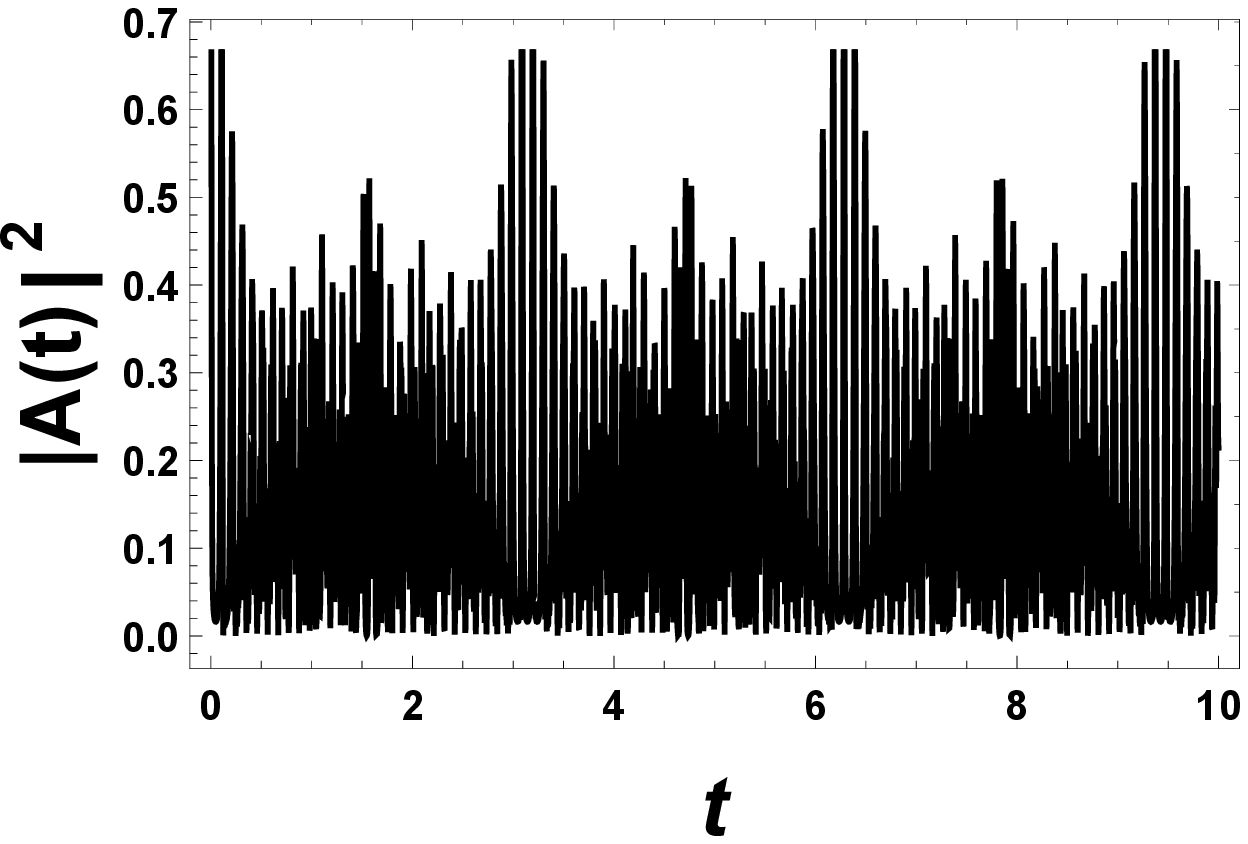}
\caption{Auto correlation function for GCS3.} \label{fidelityrat3}
\end{minipage}\hfill
\begin{minipage}{2.3in}
\centering
\includegraphics[height=2in,width=2.2in]{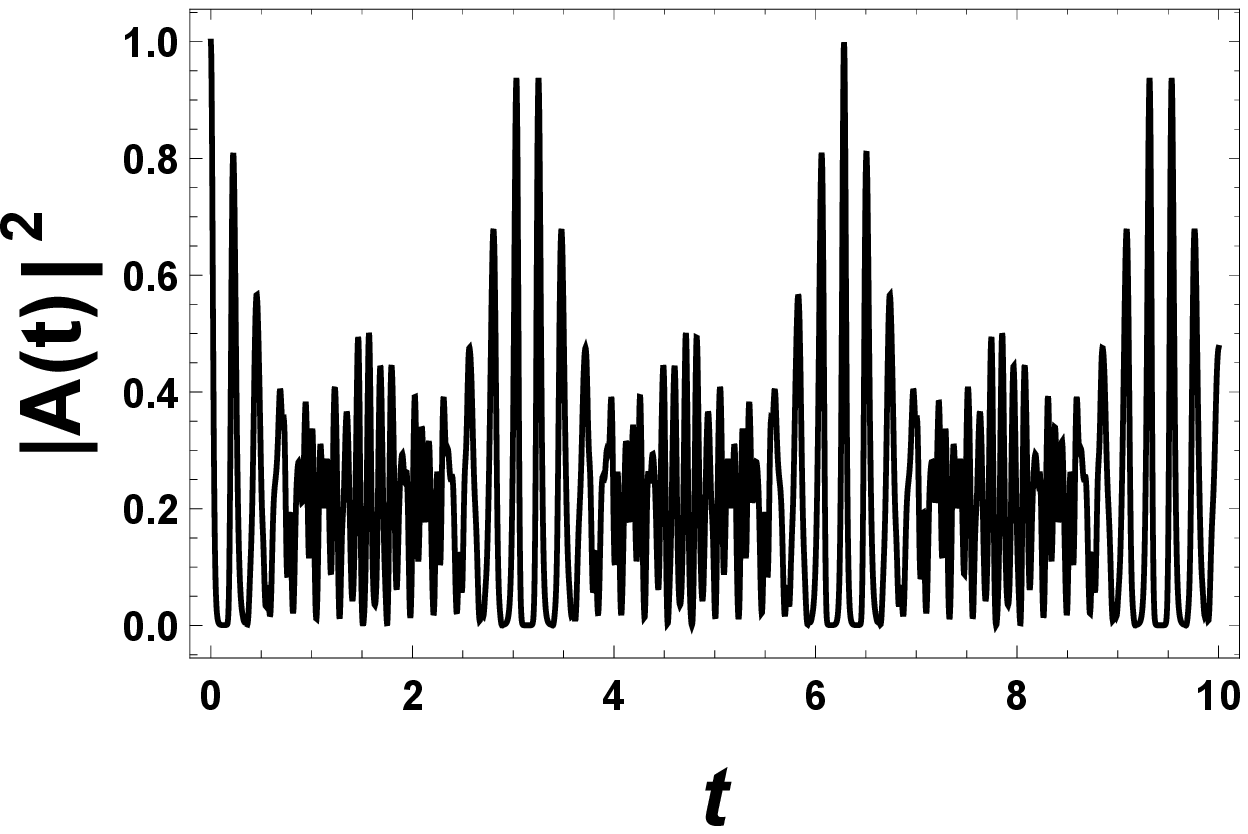}
\caption{Auto correlation function for GCS3.} \label{fidelityrat4}
\end{minipage}
\ef

\section{Statistical and Geometrical aspects}

\subsection{Statistical Property}

Various statistical properties of the GCS that have been constructed, will be presented in this section. The study of these properties gives us insight into various interesting phenomena observed in quantum optics, quantum electronics, and some foundational aspects of quantum mechanics. The classical nature of the GCS is encoded in the Poissonian distribution and any deviation from it indicates non-classical behaviour.
Note that, all the figures that follow below have been plotted for the same values of the potential parameter $\alpha$. In every figure corresponding to GCS1, 2, and 3 the blue dotted line corresponds to $\alpha = 2$, the red dot-dashed line corresponds to $\alpha = 5$, and the purple dashed line is for $\alpha = 10$ whereas for the GCS4 we have chosen $\sigma = 1.5, -1, -9$ respectively.

The deviation from classicality can be measured using the second order correlation factor $g^{(2)}$. Three possibilities exist depending on the values that $g^{(2)}$ takes. If $g^{(2)} = 1$ then the states are classical. If $g^{(2)} > 1$ it indicates bunching effect and $g^{(2)} < 1$ antibunching. In other words whenever $g^{(2)} \neq 1$, it signifies non-classsical behaviour. The second order correlation function is given by
\be \label{intcorr}		
g^{(2)} = \frac{N^{\prime \prime}(z) \ N(z)}{N^{\prime 2}(z)} \ ,
\ee
where $z = |\zeta|^2$ and the primes denote derivatives with respect to $z$. $N(z)$ is the normalization constant corresponding to the different GCS as given in equations (\ref{norm-gcs1}), (\ref{norm-gcs2}), (\ref{norm-gcs3}), and (\ref{norm-gcs4}) respectively.
\bef
\centering
\begin{minipage}{2.3in}
\centering
\includegraphics[height=2in,width=2.2in]{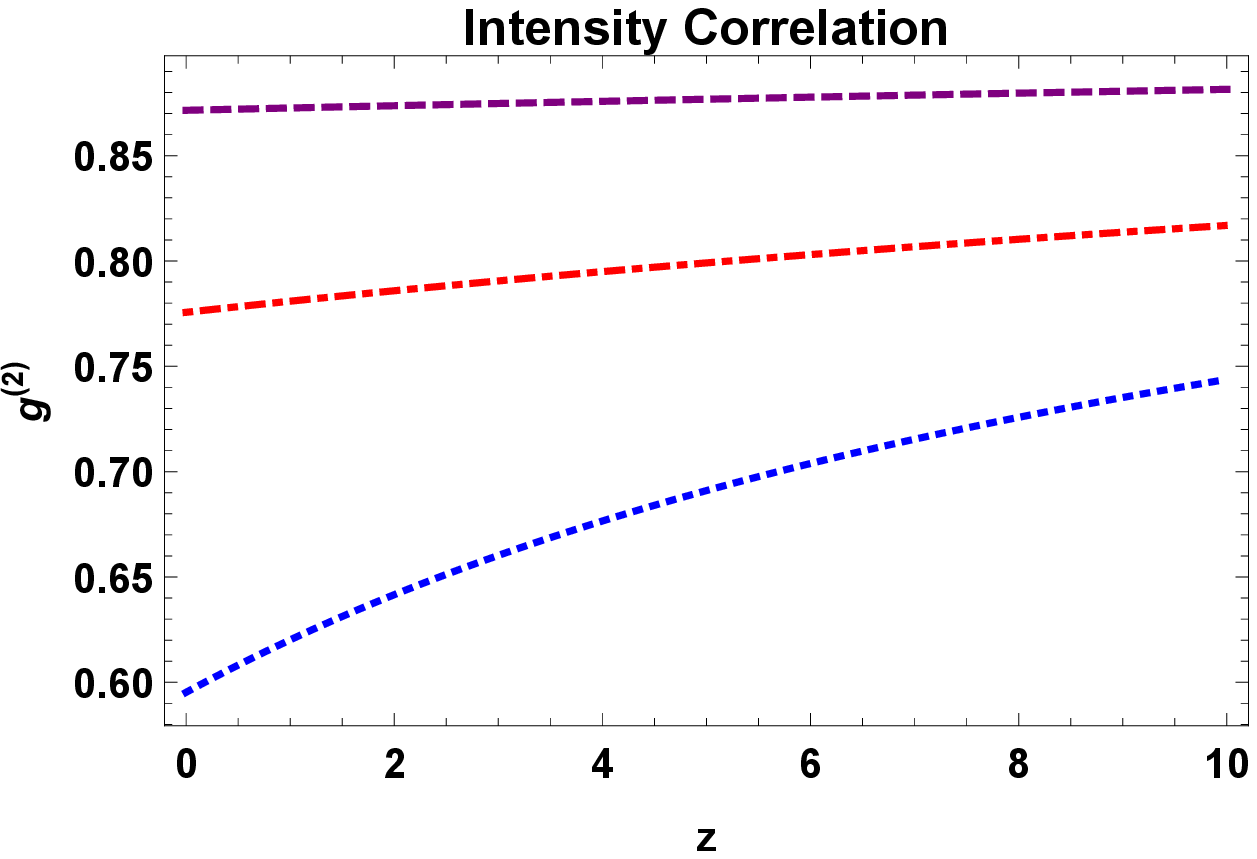}
\caption{Intensity correlation for GCS1.} \label{intcorr1}
\end{minipage}\hfill
\begin{minipage}{2.3in}
\centering
\includegraphics[height=2in,width=2.2in]{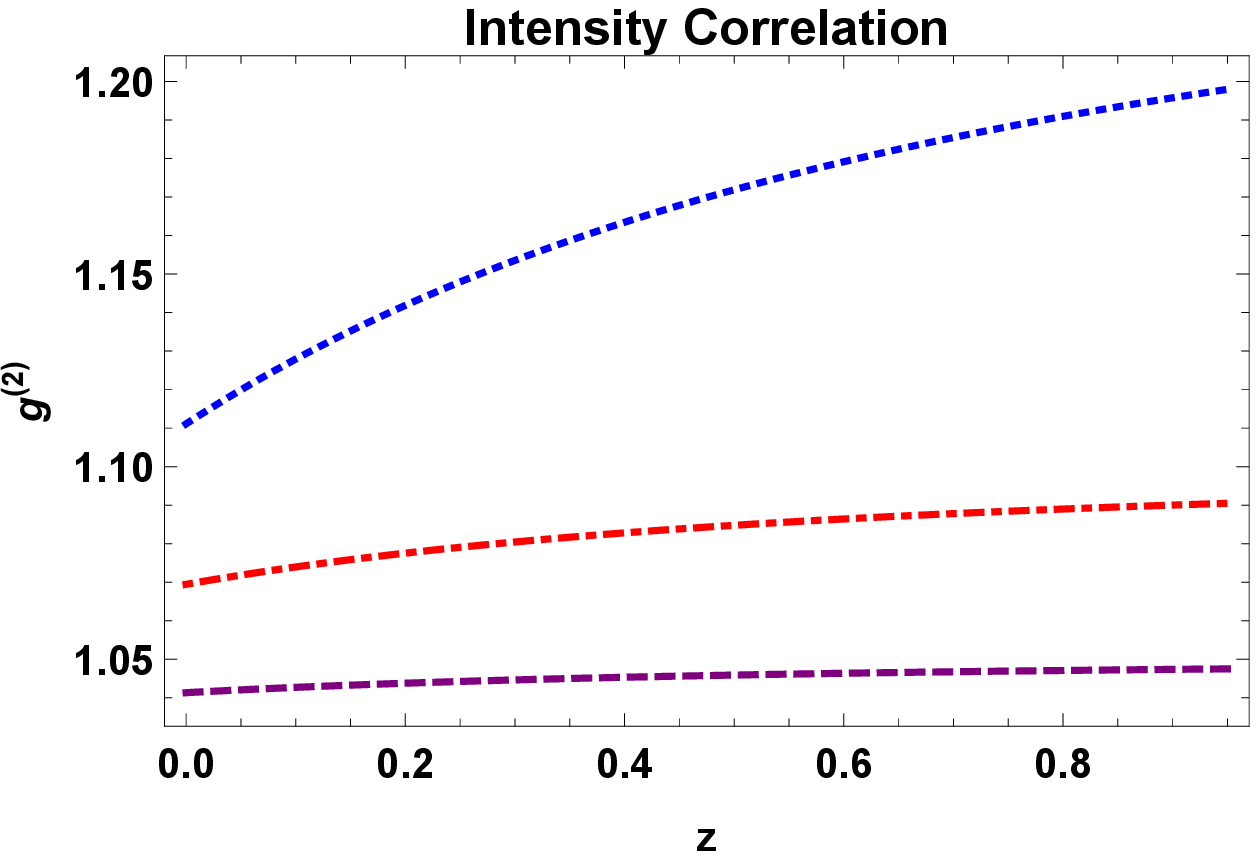}
\caption{Intensity correlation for GCS2.} \label{intcorr2}
\end{minipage}
\ef
\bef
\centering
\begin{minipage}{2.3in}
\centering
\includegraphics[height=2in,width=2.2in]{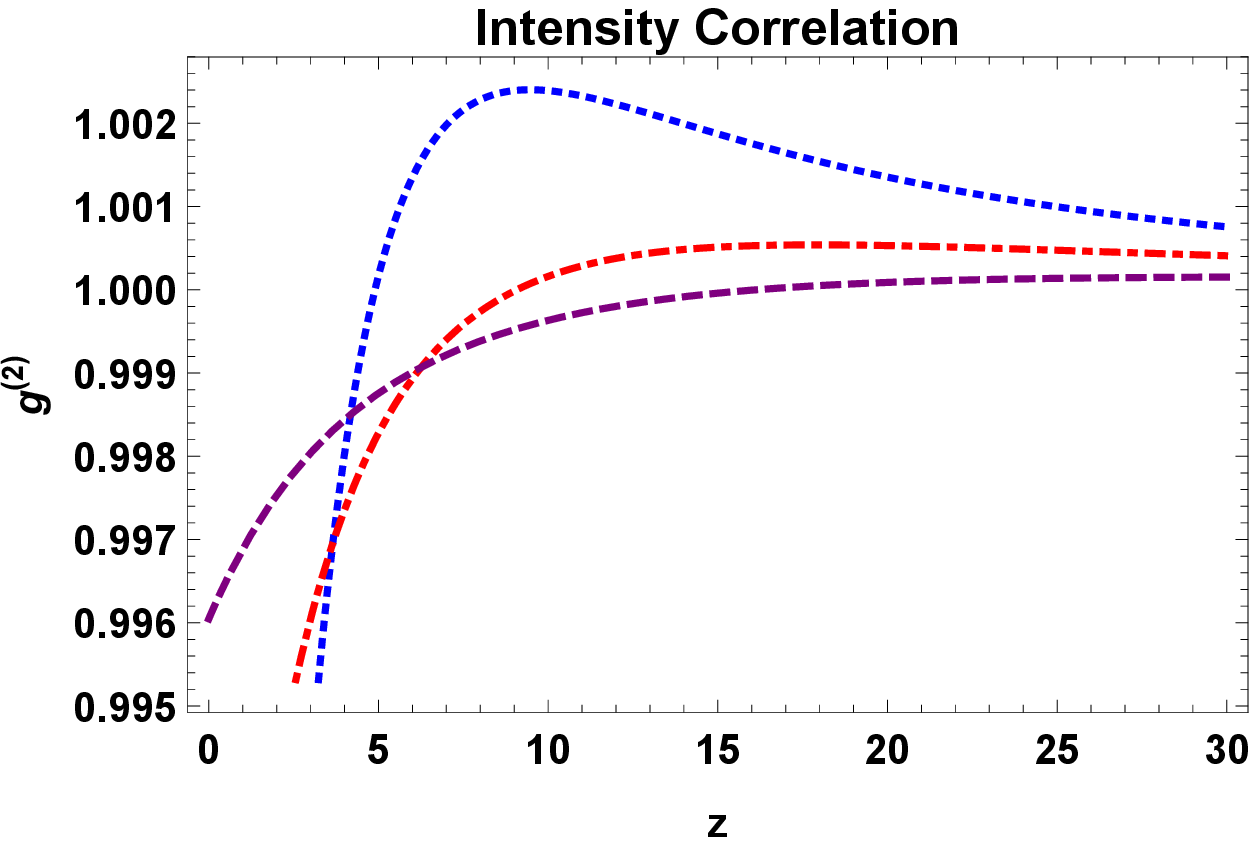}
\caption{Intensity correlation for GCS3.} \label{intcorr3}
\end{minipage}\hfill
\begin{minipage}{2.3in}
\centering
\includegraphics[height=2in,width=2.2in]{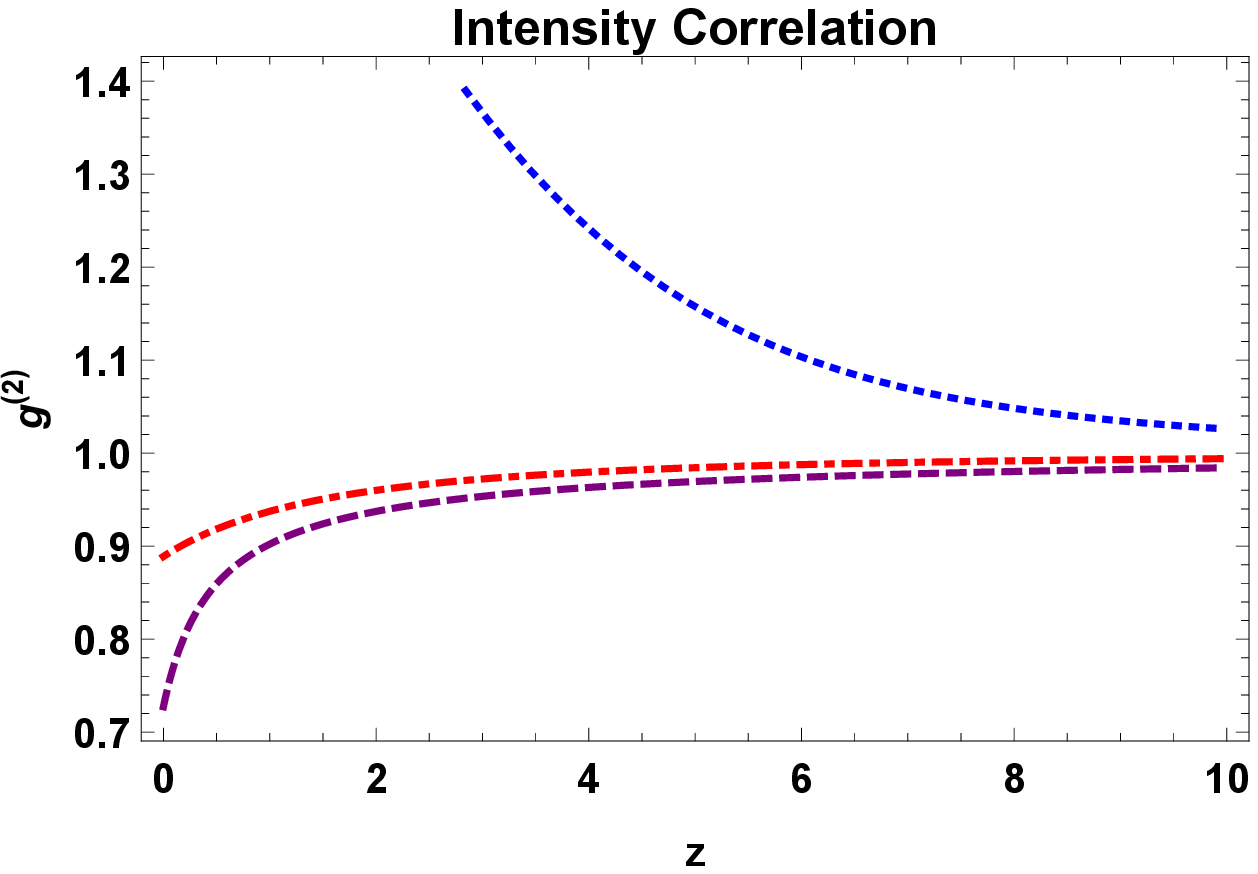}
\caption{Intensity correlation for GCS4.}\label{intcorr4}
\end{minipage}
\ef

The intensity correlation plot for the GCS1, shown in Fig. (\ref{intcorr1}), is less than one indicating antibunching effect; whereas Fig. (\ref{intcorr2}) is the plot for GCS2 which show bunching behaviour. This behaviour is seen for all values of the potential parameter considered. Figure (\ref{intcorr3}) and Figure (\ref{intcorr4}) show nonclassical for small values of $z$ but tend towards classical behaviour asymptotically.


Mandel parameter ($\mathcal{Q}$) \cite{Mandel} is another way to characterise nonclassical states. If the value of $\mathcal{Q} <  0$ the statistics is sub-Poissonian whereas, for positive values of $\mathcal{Q}$ the statistics is super-Poissonian. The Mandel parameter can be defined in terms of the normalisation constant of the CS as
\be \label{Mandelprmt}
\mathcal{Q} = z \ \left(\frac{N^{\prime \prime}(z)}{N^{\prime}(z)}-\frac{N^\prime(z)}{N(z)} \right) \ ,
\ee

\bef
\centering
\begin{minipage}{2.3in}
\centering
\includegraphics[height=2in,width=2.2in]{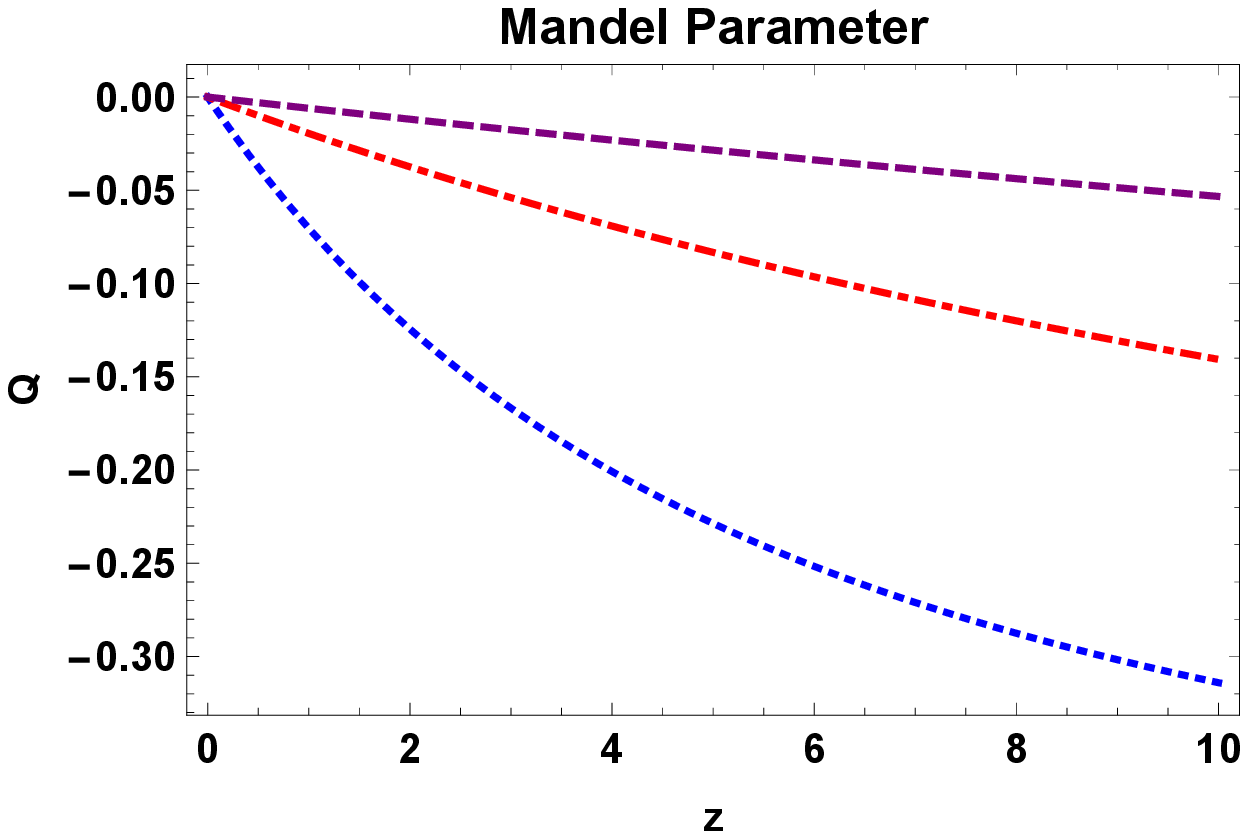}
\caption{Mandel parameter for GCS1.}\label{mandel1}
\end{minipage}\hfill
\begin{minipage}{2.3in}
\centering
\includegraphics[height=2in,width=2.2in]{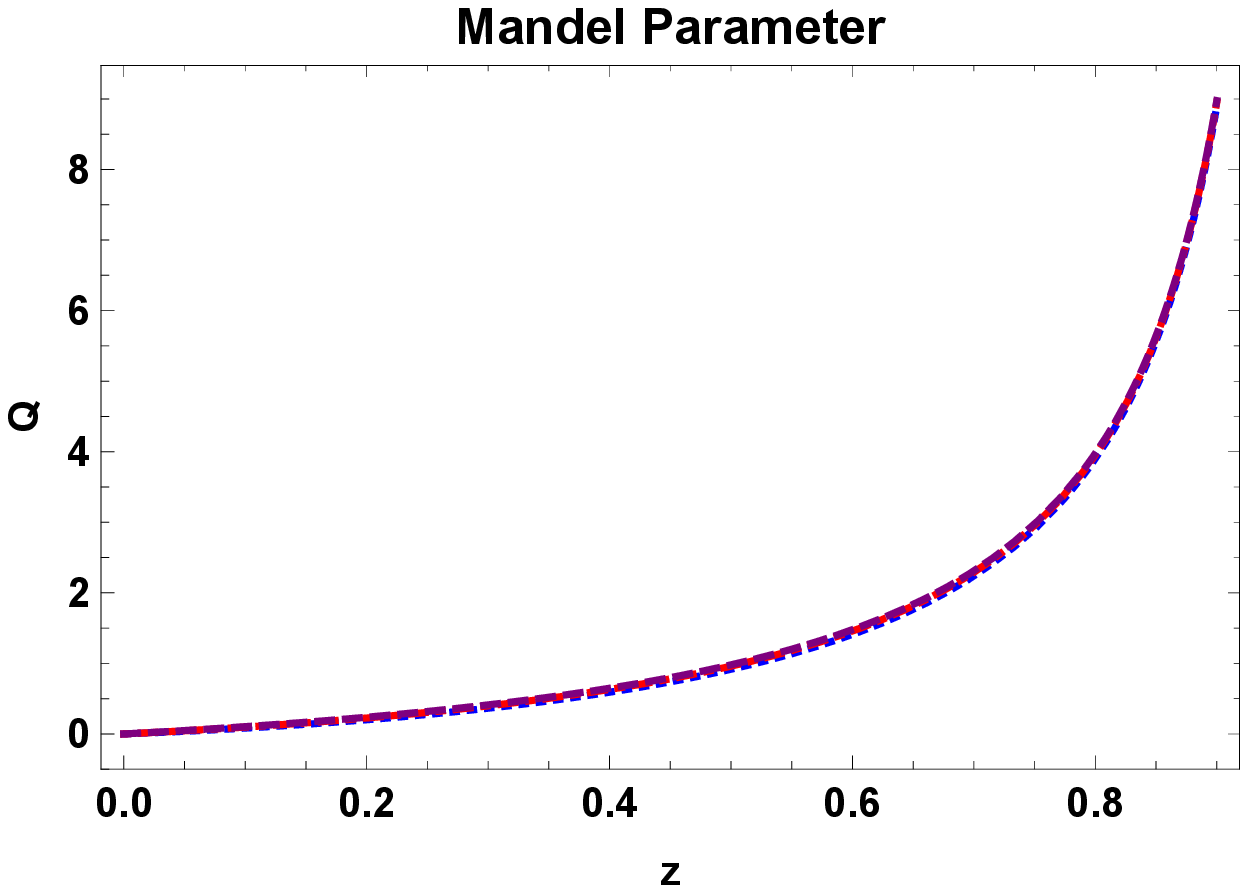}
\caption{Mandel parameter for GCS2.}\label{mandel2}
\end{minipage}
\ef
\bef
\centering
\begin{minipage}{2.3in}
\centering
\includegraphics[height=2in,width=2.2in]{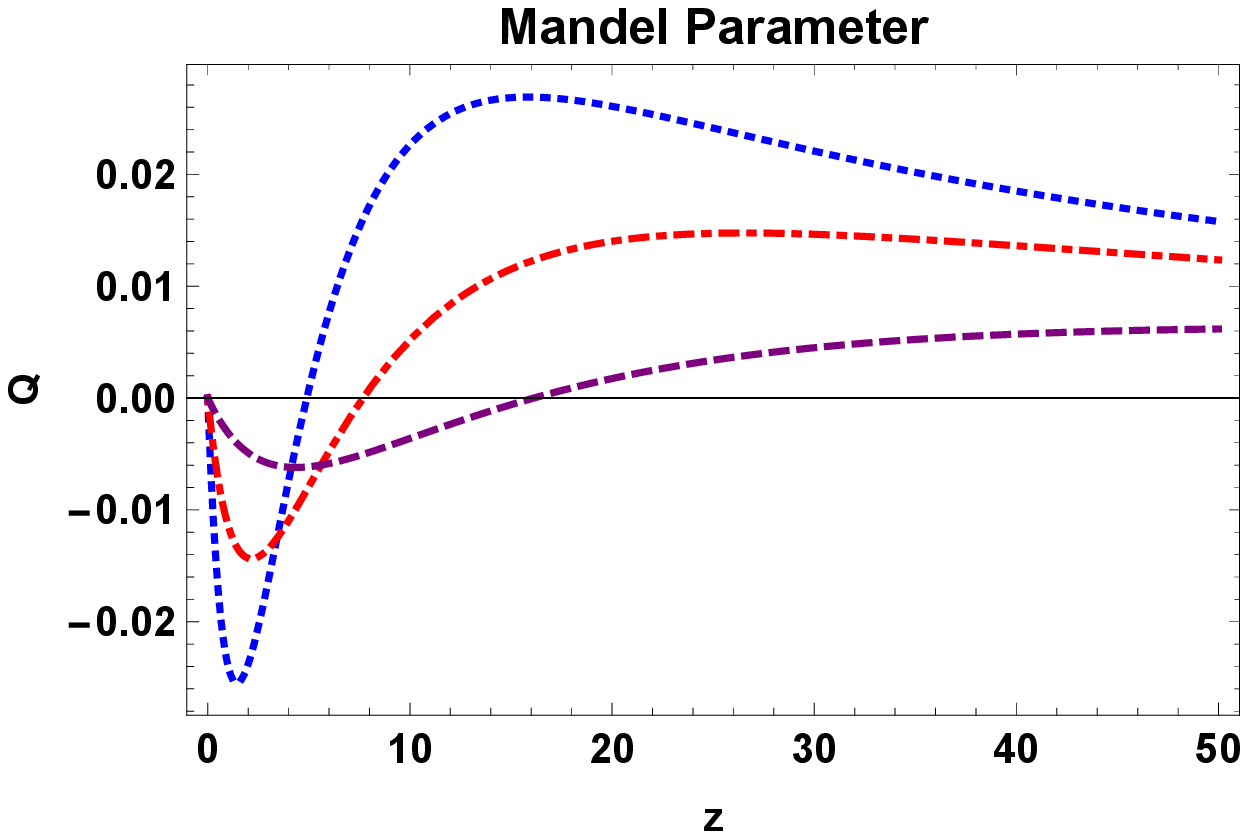}
\caption{Mandel parameter for GCS3.}\label{mandel3}
\end{minipage}\hfill
\begin{minipage}{2.3in}
\centering
\includegraphics[height=2in,width=2.2in]{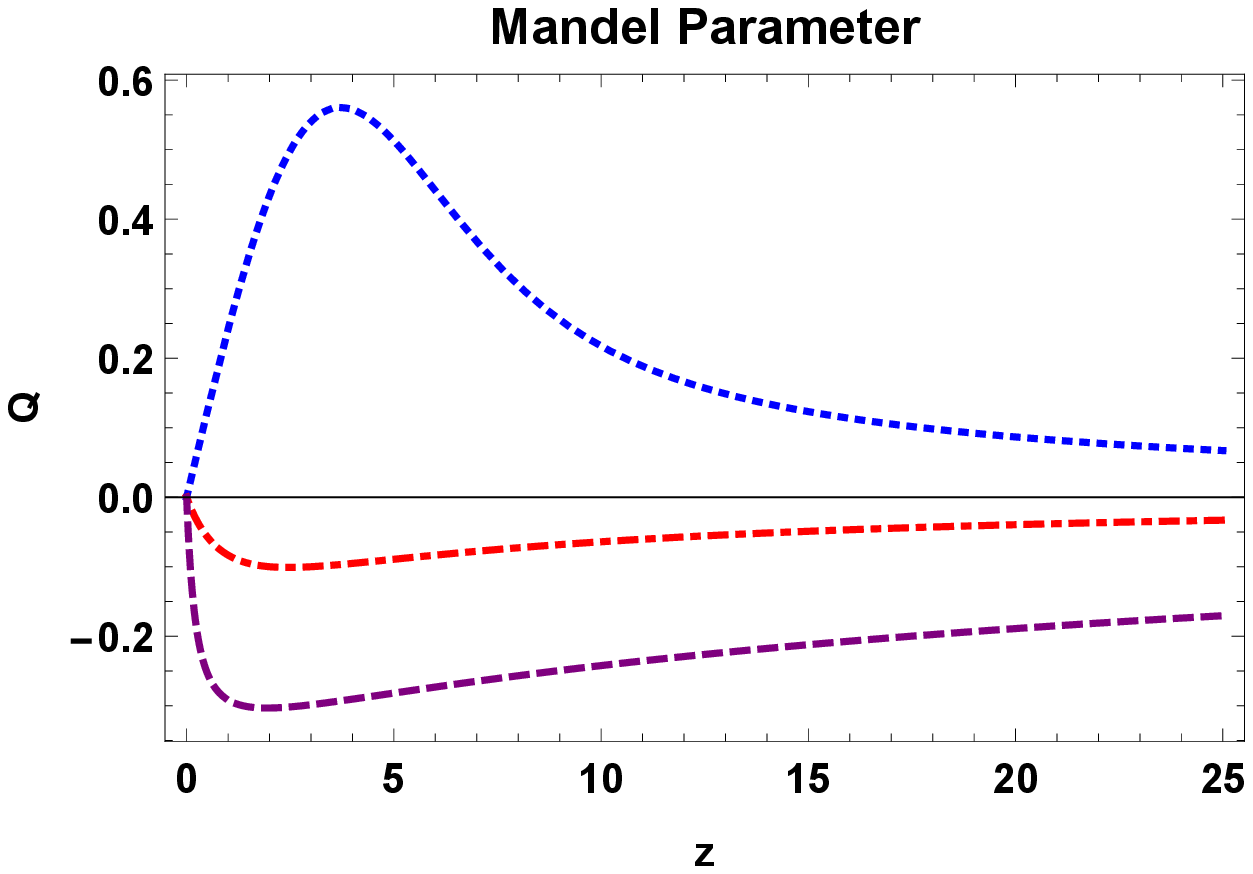}
\caption{Mandel parameter for GCS4.}\label{mandel4}
\end{minipage}
\ef

The plots of the Mandel parameter for GCS1 and GCS2 are shown in Figures (\ref{mandel1}) and (\ref{mandel2}) respectively.  The former shows sub-Poissonian behaviour while the latter super-Poissonian. Interestingly the GCS2 has the same $Q$ values for all values of $\alpha$ and $z$. Figures (\ref{mandel3}) and (\ref{mandel4}) coresponding to GCS3 and GCS4 respectively, yield a similar information as the intensity correlation graphs i.e., the CS approach classicality for large values of $z$.


In what follows, we plot the weighting distributions, defined as $P = |\langle \zeta; a_r | \zeta; a_r  \rangle|^2$, for the GCS 1-4 in Figures (\ref{photdist1}),  (\ref{photdist2}),  (\ref{photdist3}), and  (\ref{photdist4}) respectively.
%
\bef
\centering
\begin{minipage}{2.3in}
\centering
\includegraphics[height=2in,width=2.2in]{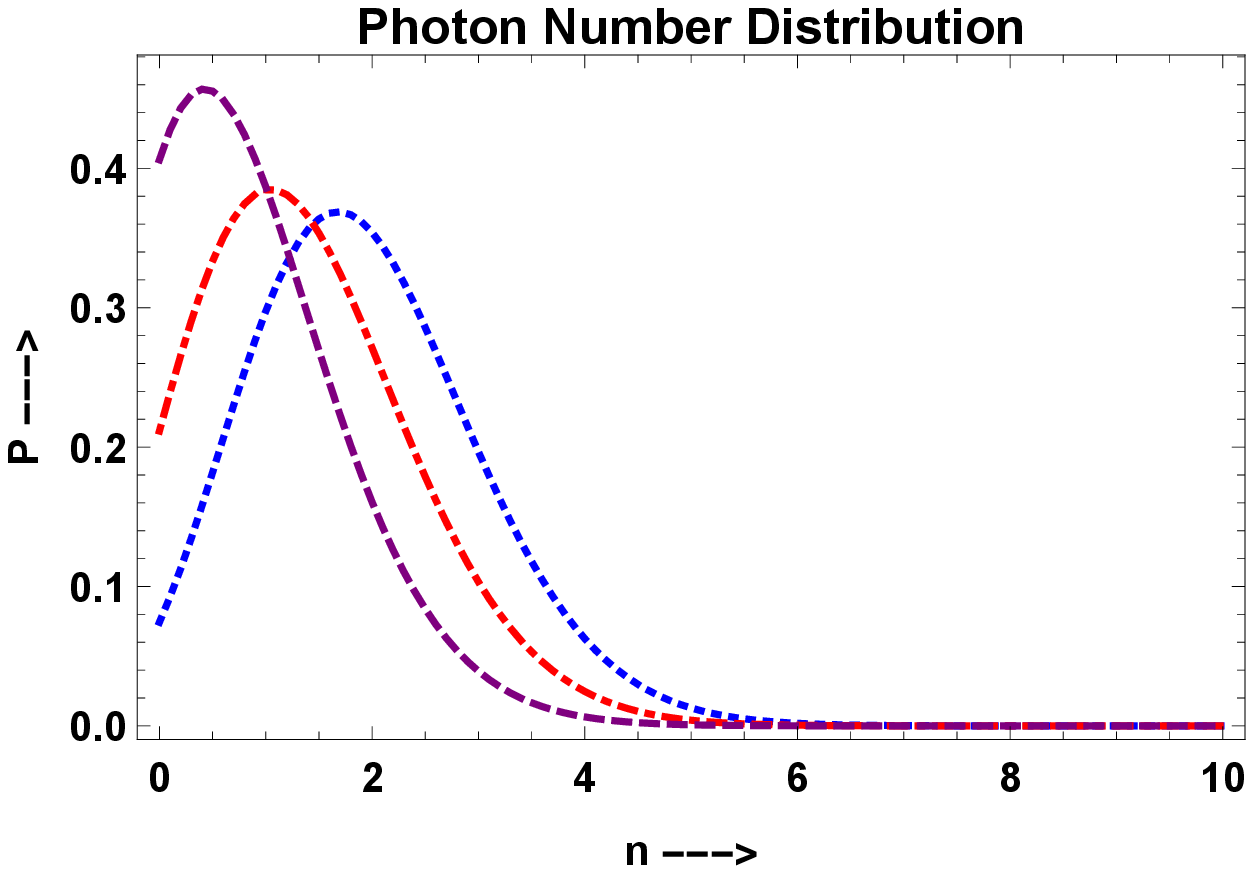}
\caption{Photon distribution GCS1.}\label{photdist1}
\end{minipage}\hfill
\begin{minipage}{2.3in}
\centering
\includegraphics[height=2in,width=2.2in]{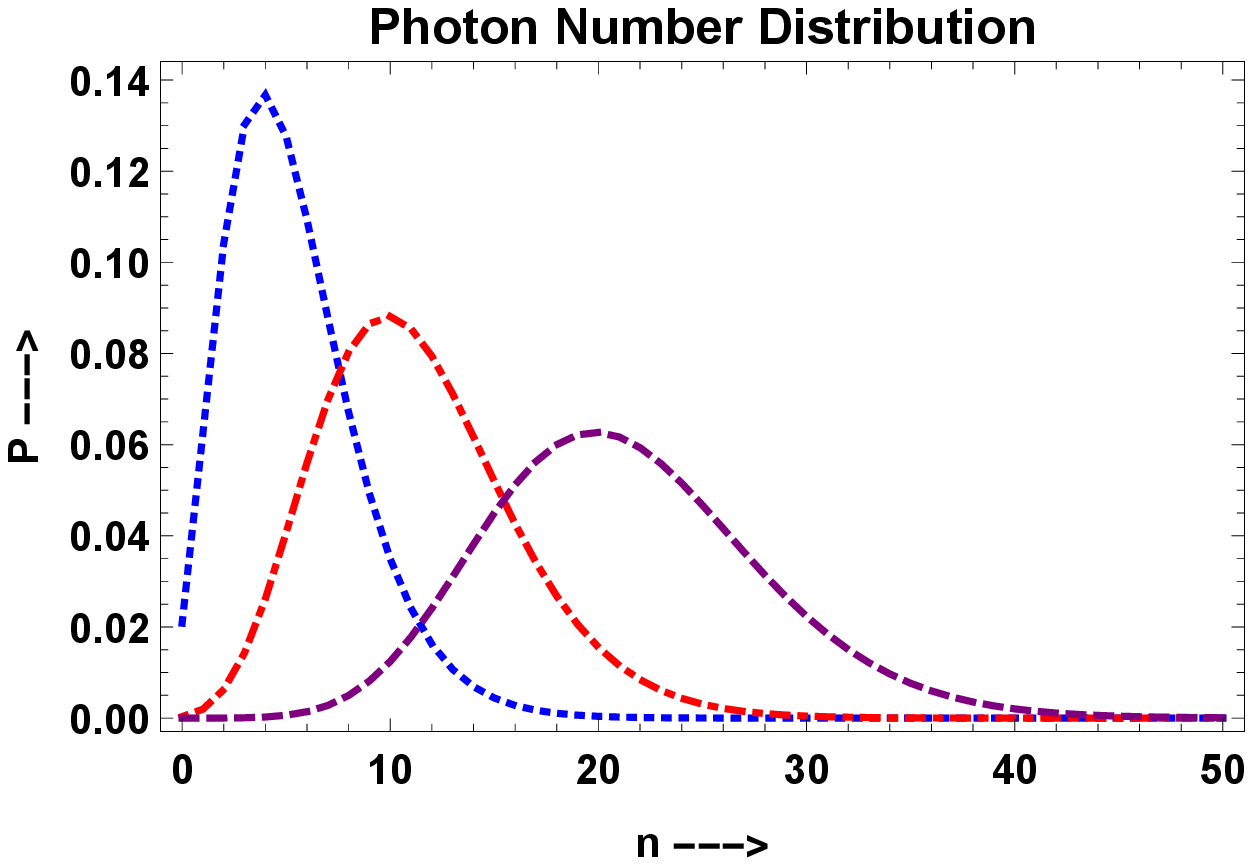}
\caption{Photon distribution GCS2.} \label{photdist2}
\end{minipage}
\ef
\bef
\centering
\begin{minipage}{2.3in}
\centering
\includegraphics[height=2in,width=2.2in]{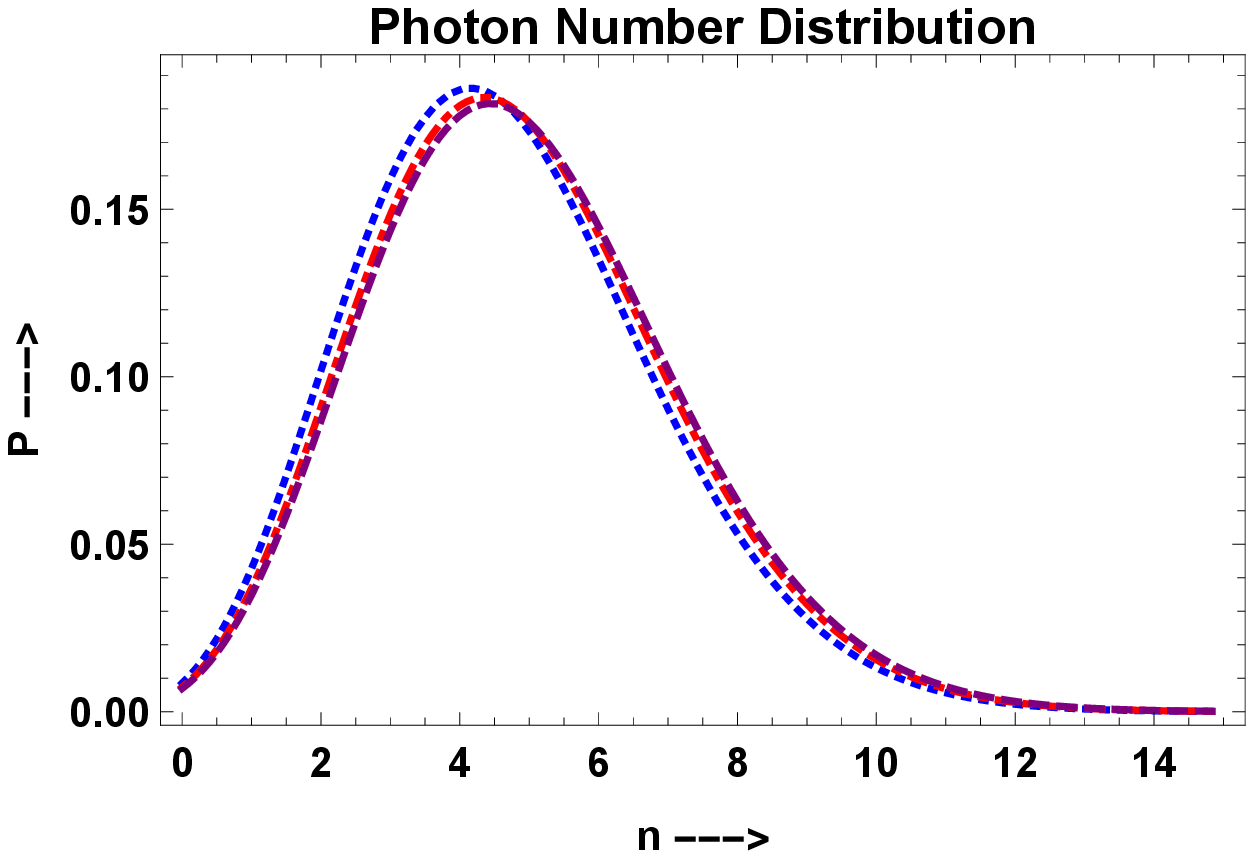}
\caption{Photon distribution GCS3.}\label{photdist3}
\end{minipage}\hfill
\begin{minipage}{2.3in}
\centering
\includegraphics[height=2in,width=2.2in]{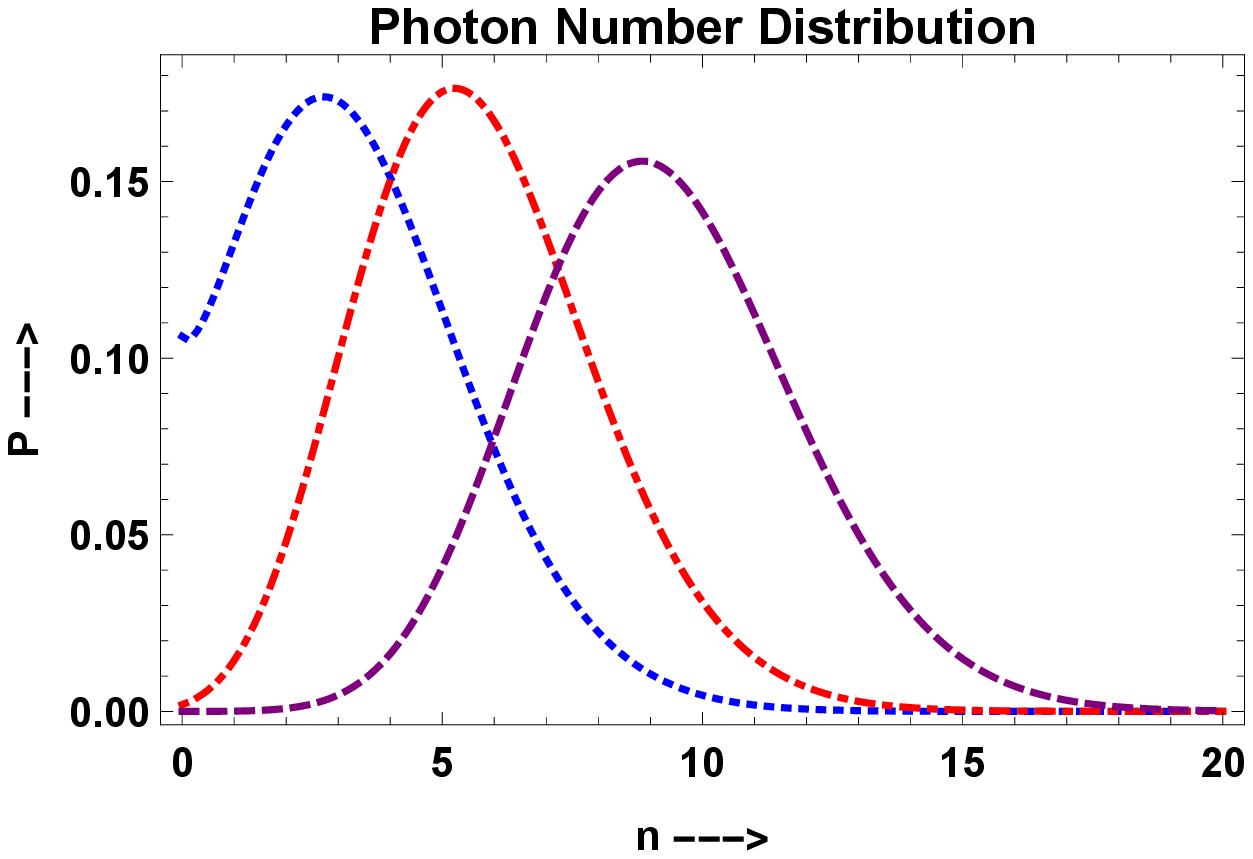}
\caption{Photon distribution GCS4.}\label{photdist4}
\end{minipage}
\ef
Similarly the mean photon number which can be defined as
\be \label{mphotonumb}
\mathcal{N} = z \ \frac{N^\prime(z)}{N(z)} \ ,
\ee
have been plotted in Figures (\ref{mphotnumb1}), (\ref{mphotnumb2}), (\ref{mphotnumb3}), and (\ref{mphotnumb4}).
\bef
\centering
\begin{minipage}{2.3in}
\centering
\includegraphics[height=2in,width=2.2in]{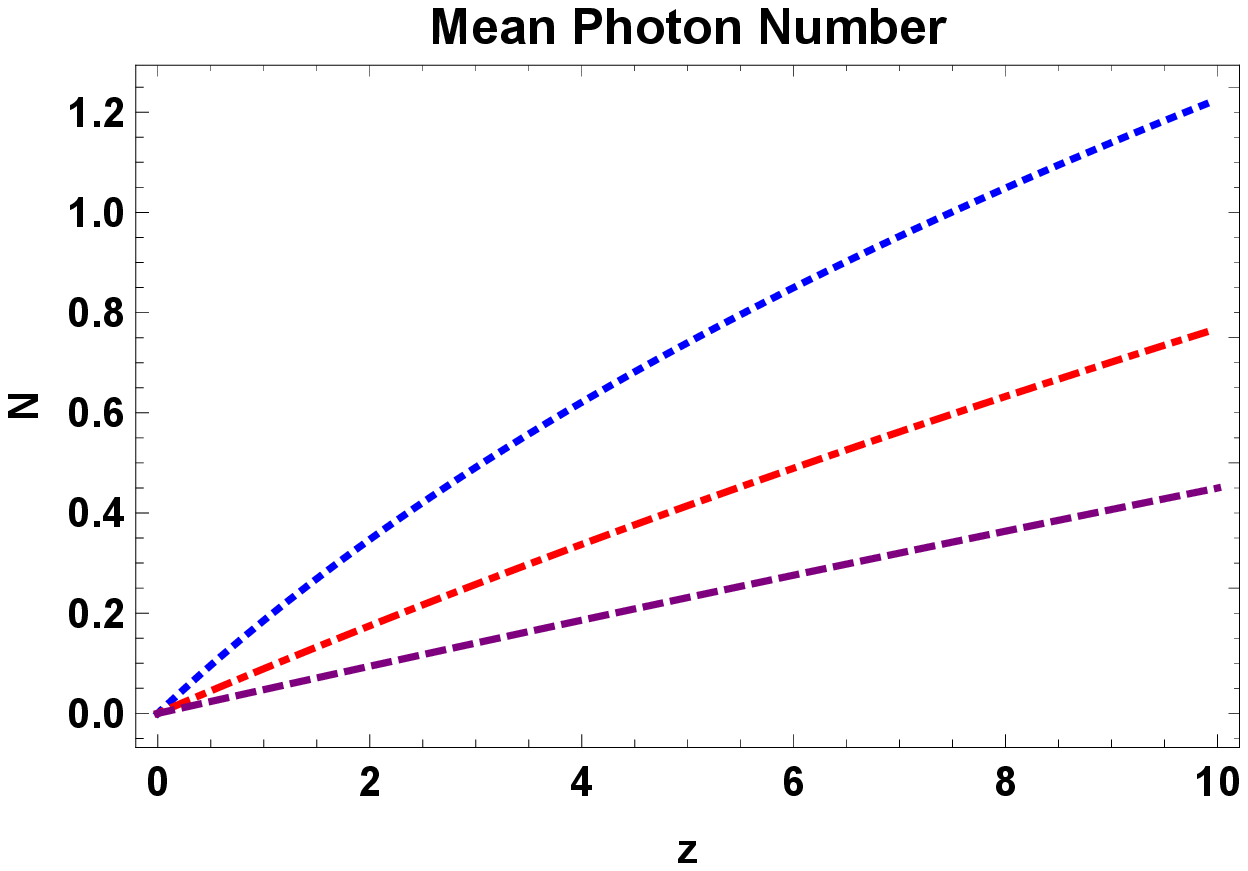}
\caption{Mean photon number for GCS1.} \label{mphotnumb1}
\end{minipage}\hfill
\begin{minipage}{2.3in}
\centering
\includegraphics[height=2in,width=2.2in]{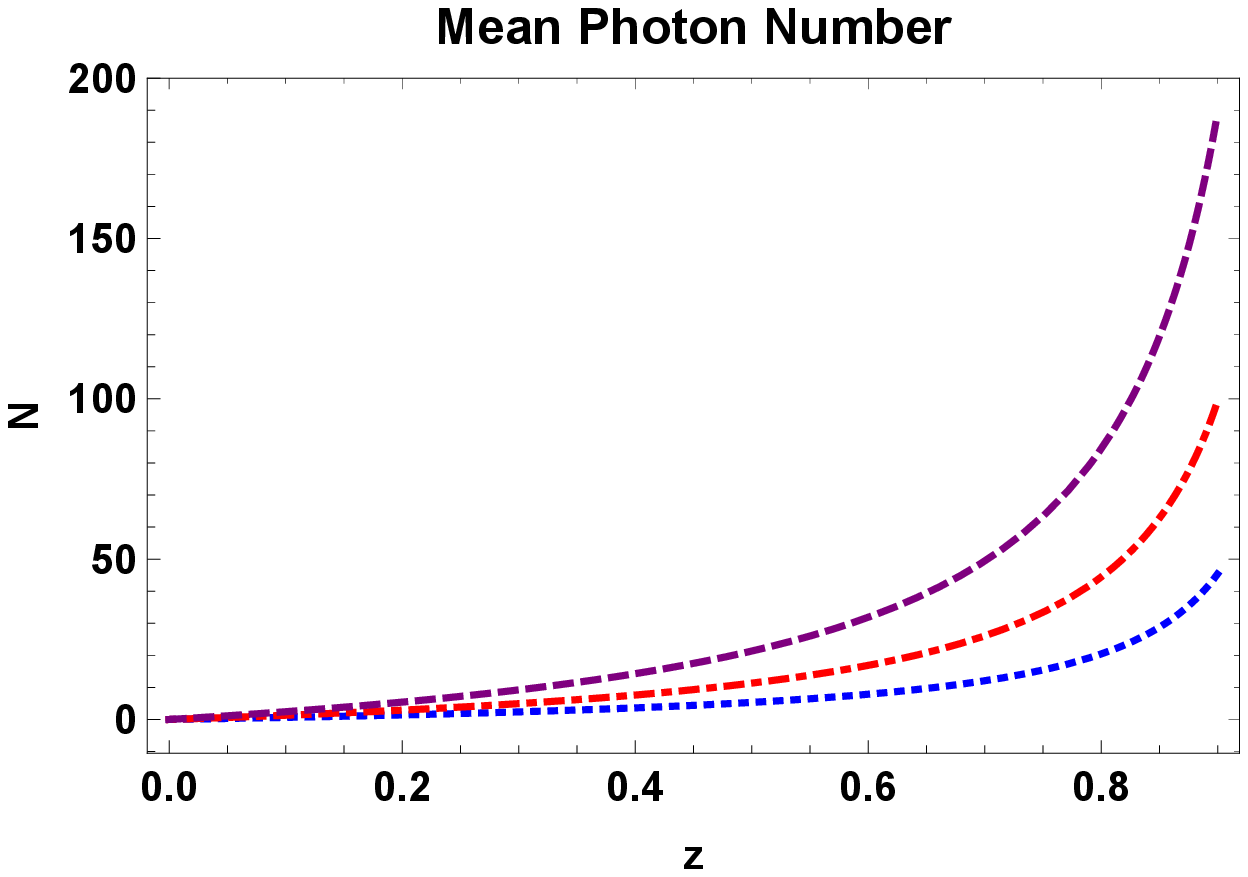}
\caption{Mean photon number for GCS2.} \label{mphotnumb2}
\end{minipage}
\ef
\bef
\centering
\begin{minipage}{2.3in}
\centering
\includegraphics[height=2in,width=2.2in]{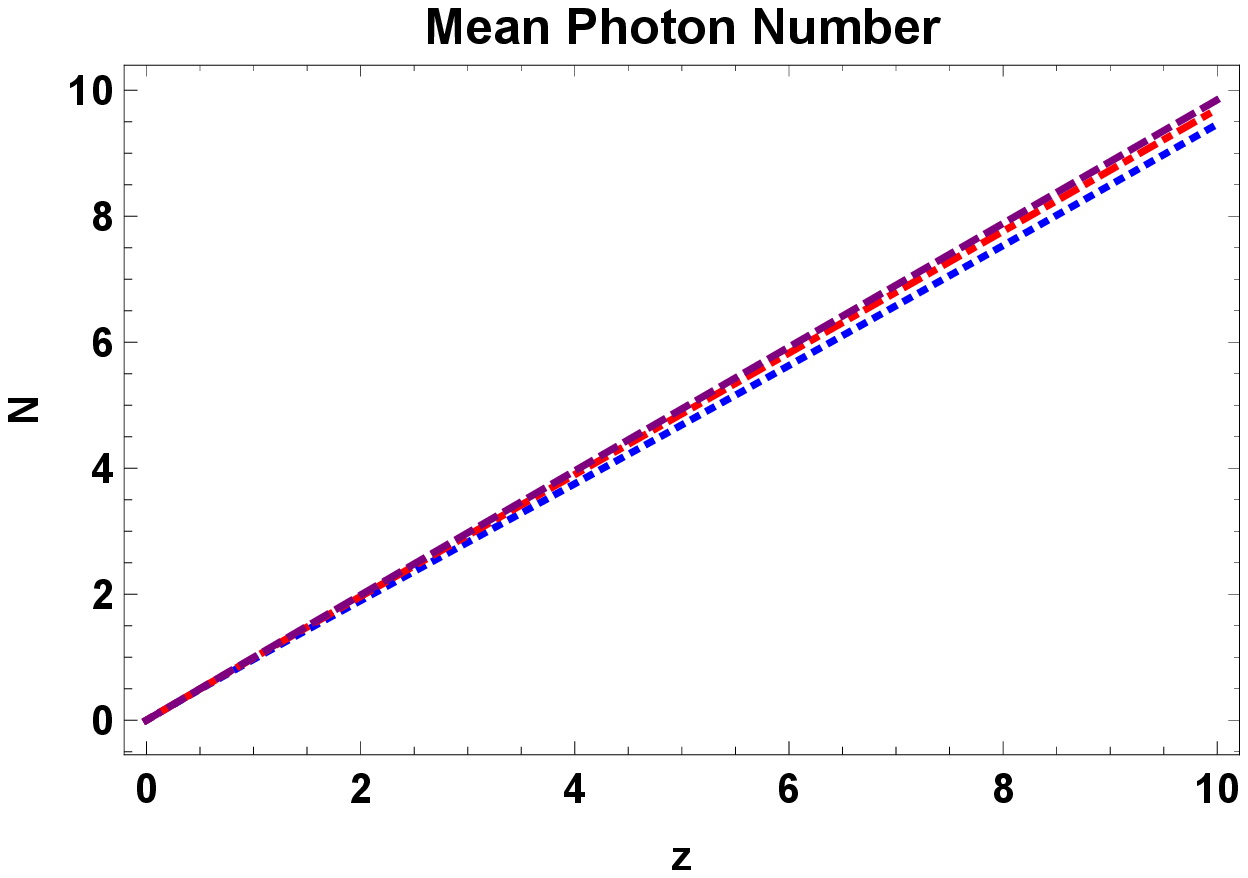}
\caption{Case 3 mean photon number.} \label{mphotnumb3}
\end{minipage}\hfill
\begin{minipage}{2.3in}
\centering
\includegraphics[height=2in,width=2.2in]{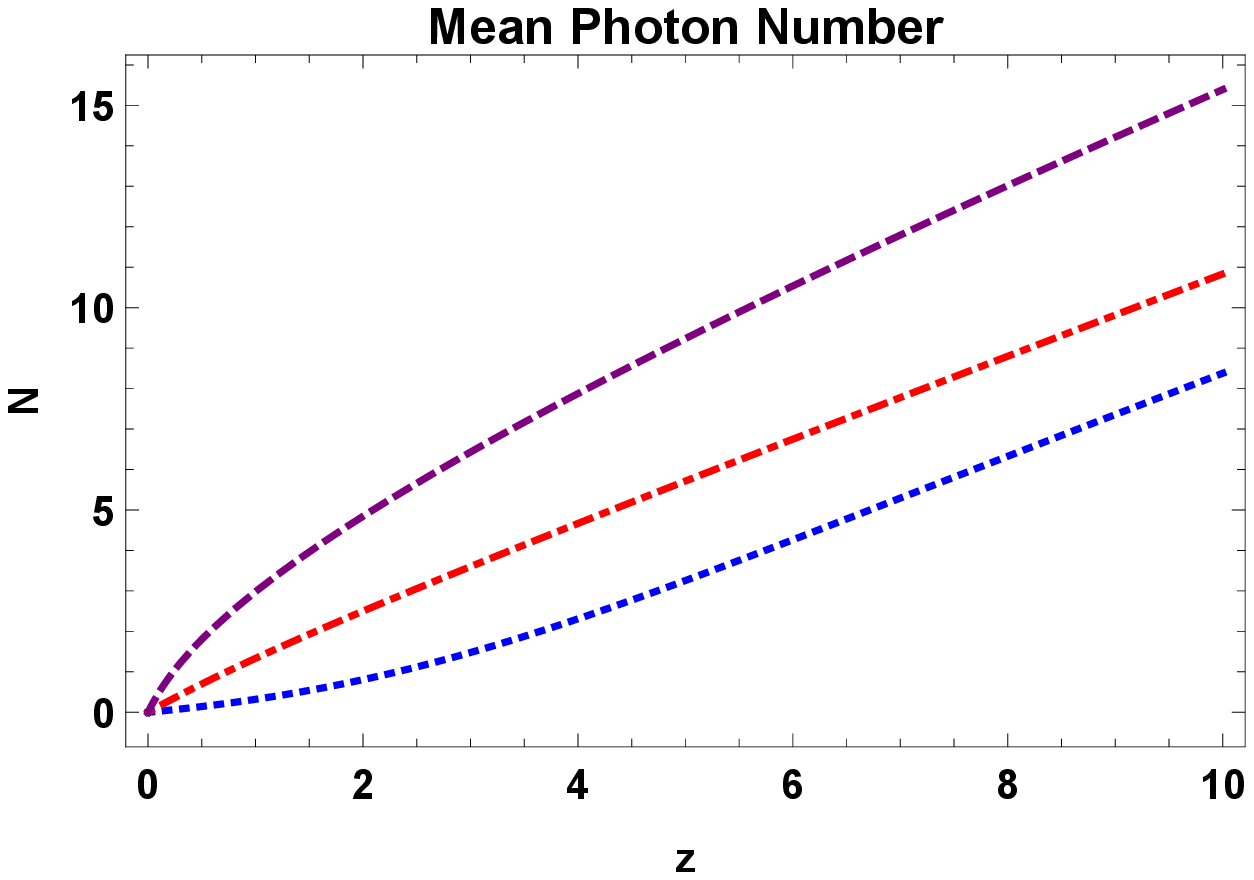}
\caption{Case 4 mean photon number.} \label{mphotnumb4}
\end{minipage}
\ef


\subsection{Geometrical Property}

The GCS constructed above have not only interesting statistical behaviour they also have fascinating geometric structure. Perelomov \cite{Perelomov} and Gilmore's \cite{Gilmore1,Gilmore2} construction of the CS has a group manifold interpretation. This group manifold exposition of the CS runs into trouble when the Barut-Girardello CS is constructed. This is because it is not based on the group theoretical aspect. This does not mean that the situation is hopeless as far as geometry is concerned. One can resort to something called the metric factor $\omega$, first introduced by Klauder et. al. \cite{Klauder}, and still gain an understanding of the geometric nature of the GCS. 

It is a well known fact that CS provide a mapping between the set of complex numbers and the rays in Hilbert space. This projective Hilbert space can be endowed with Fubini-Study metric. This metric can be written via the line element in terms of the CS as $ds^2:= || d |\zeta \rangle ||^2 - |\langle \zeta | d | \zeta \rangle |^2$ \cite{Klauder}. 

The metric factor surprisingly can be expressed entirely in terms of the normalisation constant of the CS:
\be \label{metricfact}
\omega(z) = \frac{N^\prime(z)}{N(z)} + z \ \left(\frac{N^{\prime \prime}(z)}{N(z)}-\frac{N^{\prime 2}(z)}{N^2(z)} \right).
\ee
If $\omega(z) = 1$ describes a flat two dimensional surface and when $\omega(z) \neq 1$ it is nonflat. For more details we refer to \cite{Klauder}.

The graphs corresponding to the GCS 1-4 are shown in Figures (\ref{metric1}), (\ref{metric2}), (\ref{metric3}), and (\ref{metric4}). From the Fig (\ref{metric1}) which corresponds to GCS1, we conclude that the surface of the CS is non-flat. Fig (\ref{metric2}), which happens to be a Perelomov CS, we can see that it starts out as a flat surface and becomes non flat as $z$ increases. In figures (\ref{metric3}) and (\ref{metric4}) the CS surfaces become flat asymptotically.

\bef
\centering
\begin{minipage}{2.3in}
\centering
\includegraphics[height=2in,width=2.2in]{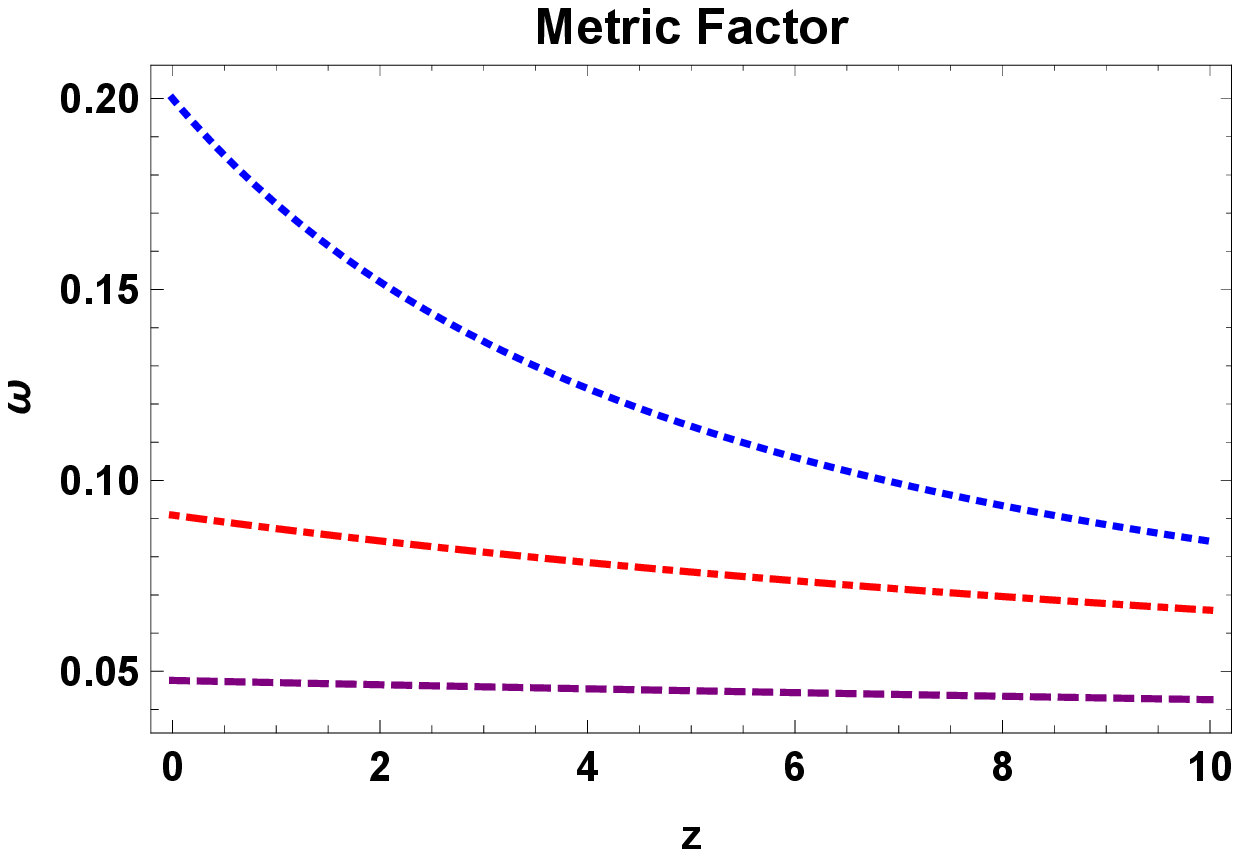}
\caption{Metric factor for GCS1.} \label{metric1}
\end{minipage}\hfill
\begin{minipage}{2.3in}
\centering
\includegraphics[height=2in,width=2.2in]{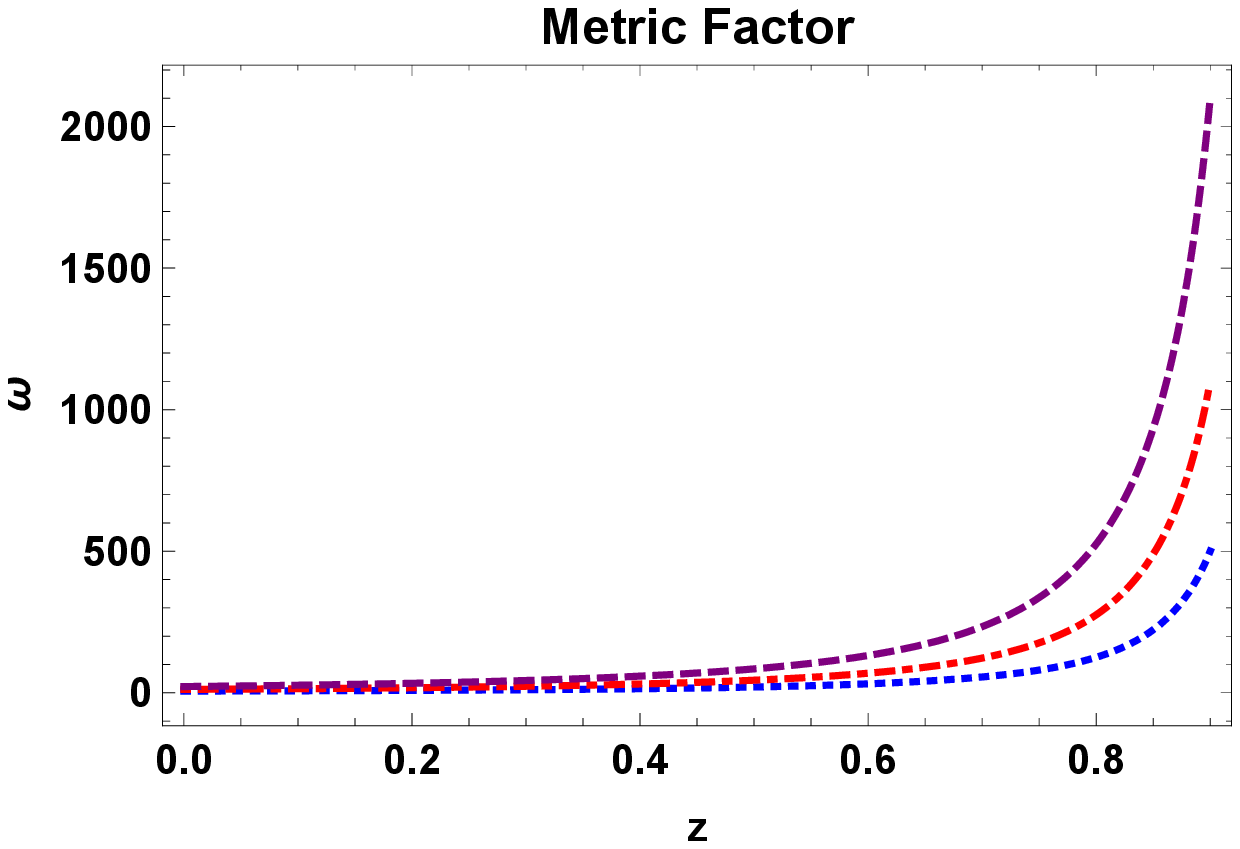}
\caption{Metric factor for GCS2.} \label{metric2}
\end{minipage}
\ef
\bef
\centering
\begin{minipage}{2.3in}
\centering
\includegraphics[height=2in,width=2.2in]{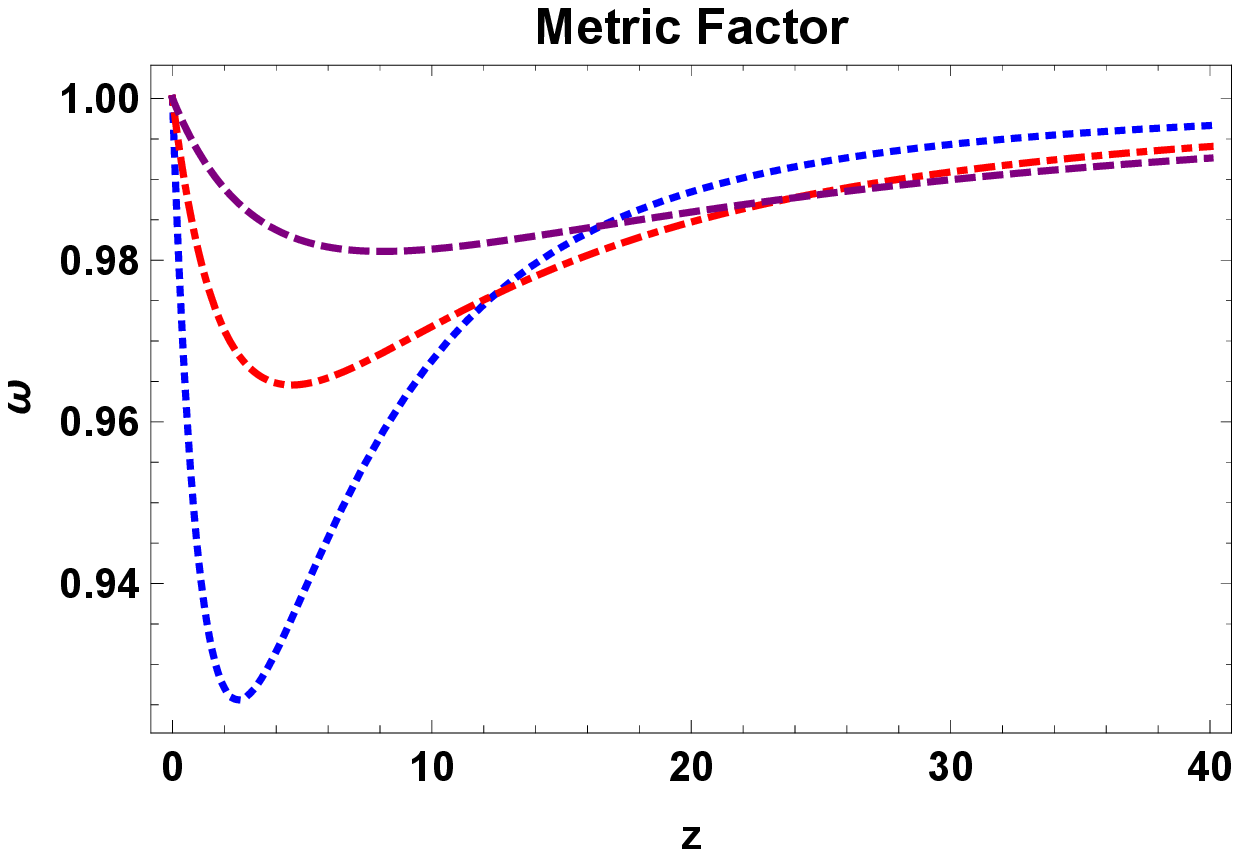}
\caption{Metric factor for GCS3.} \label{metric3}
\end{minipage}\hfill
\begin{minipage}{2.3in}
\centering
\includegraphics[height=2in,width=2.2in]{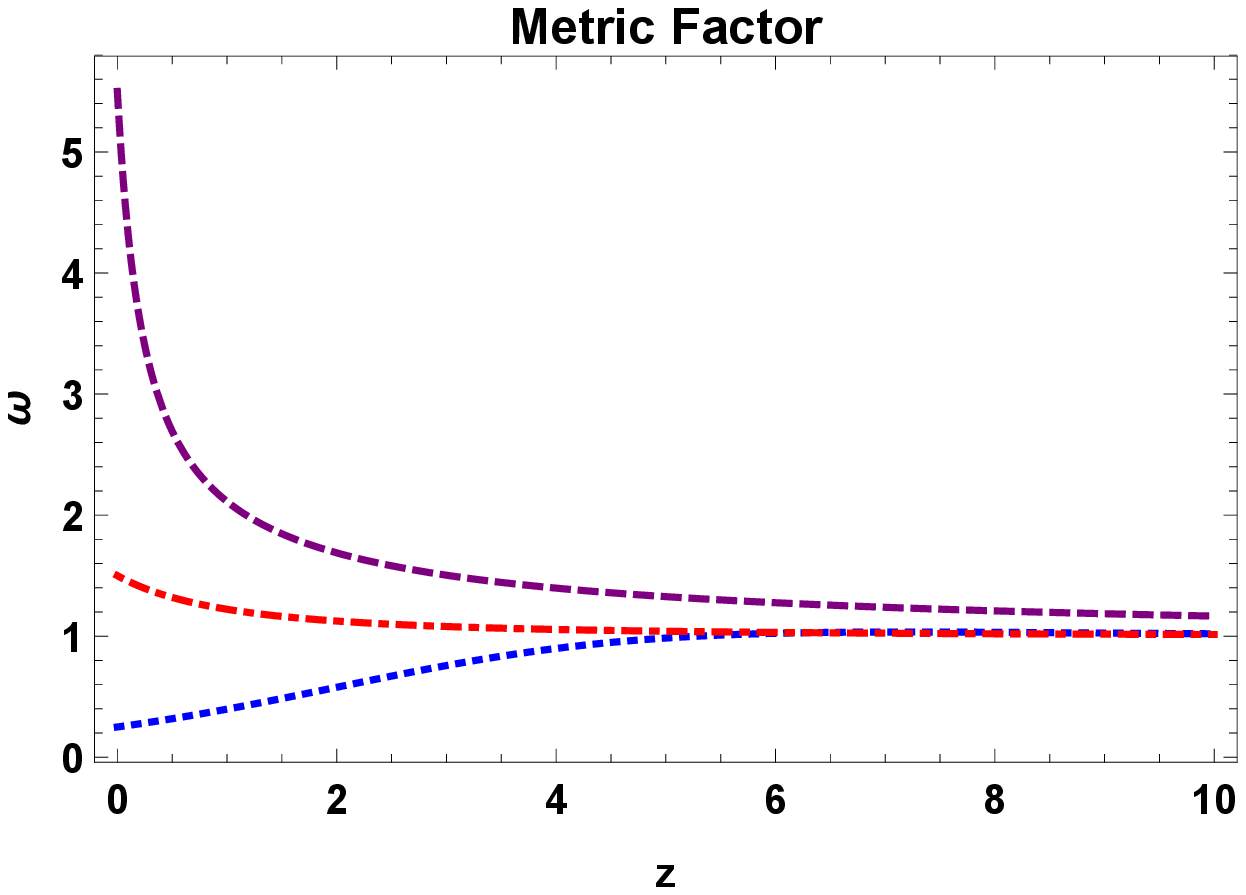}
\caption{Metric factor for GCS4.} \label{metric4}
\end{minipage}
\ef

\section{Conclusions}

In this paper we have constructed the GCS for the rational Scarf-I and the Scarf-I potential. We  looked at their quantum carpet structures and have made a comparative study. It is noticed that the quantum carpets for the rational potential show intricate sub-structures as compared to the carpets of the conventional potential. We were able to trace this to the increased asymmetry in the potential, due to the presence of the rational terms in the exceptional Scarf-I potential. This shows that these new potentials have interesting revival dynamics. The auto-correlation studies also confirm the same. The statistical aspects have been studied via the intensity correlation, Mandel parameter, mean photon number, and the probability distribution. The geometrical aspects were investigated using the metric factor. A more detailed analysis along the lines of \cite{Lane} needs to be done and we hope to return to it in future.

\vskip0.5cm
\noindent \textbf{Acknowledgements}
\vskip0.2cm
\noindent TS is supported by SERB, India Grant: ECR/2015/000081. SSR is supported by SERB, India Grant: EMR/2016/005002.

\end{document}